\RequirePackage{fix-cm}
\documentclass[epjc3]{svjour3}  

\RequirePackage{graphicx,amssymb,amsfonts,amsmath,geometry,hyperref,color,lipsum,soul}
\usepackage[english]{babel} 
\usepackage[T1]{fontenc}
\usepackage{float}
\usepackage{pdflscape}
\usepackage{multirow}
\journalname{Eur. Phys. J. C}

\begin{document}
\title{Modified holographic Ricci interacting dark energy models: dynamical system analysis and bayesian comparison.}
\author{Antonella Cid M.\thanksref{e1,addr1, addr1b}
        \and  Israel Obreque\thanksref{e2,addr2}
}
\thankstext{e1}{e-mail: acidm@ubiobio.cl}
\thankstext{e2}{e-mail: israel.obreque@usach.cl}

\institute{Departamento de F\'isica, Universidad del B\'io-B\'io, Casilla 5-C, Concepci\'on, Chile. \label{addr1} \and
Centro de Ciencias Exactas, Universidad del B\'io-B\'io, Casilla 447, Chill\'an, Chile.
\label{addr1b} \and Departamento de F\'isica, Universidad de Santiago de Chile, Casilla 307, Santiago, Chile \label{addr2}}

\date{Received: date / Accepted: date} 
\maketitle

\begin{abstract}

We perform a dynamical system analysis and a Bayesian model selection for a new set of interacting scenarios in the framework of modified holographic Ricci dark energy models (MHR-IDE). The dynamical analysis shows a modified radiation epoch and a late-time attractor corresponding to dark energy. We use a combination of background data such as type Ia supernovae, cosmic chronometers, cosmic microwave background, and baryon acoustic oscillations measurements. We find evidence against all the MHR-IDE scenarios studied with respect to $\Lambda$CDM, when the full joint analysis is considered.
\end{abstract}


\section{Introduction}\label{sec1}

Our universe is experiencing accelerated expansion \cite{Weinberg:2013agg}. This acceleration is attributed to dark energy, represented by a cosmological constant $\Lambda$ in the standard cosmological model. While the standard cosmological model, also known as $\Lambda$CDM, is a suitable model for the current expansion and the phases observed in the universe's evolution, it is plagued by several issues. These include the cosmological constant problem \cite{Weinberg:1988cp,Weinberg:2000yb}, the coincidence problem \cite{Chimento:2003iea,Guo:2004vg,Pavon:2005yx}, and discrepancies in the values of the Hubble parameter obtained from local measurements and those inferred from Planck's data \cite{Riess:2021jrx}.

Over the past two decades, numerous models of dark energy (DE) have emerged to account for the current accelerated expansion of the universe (see for example Refs. \cite{Copeland:2006wr}, \cite{Yoo:2012ug}, \cite{Wang:2016och}, and \cite{Bahamonde:2017ize} and references therein). 
Holographic dark energy models (HDE), in particular, are built upon the holographic principle, as initially proposed by 't Hooft \cite{tHooft:1993dmi}. In this framework, the authors of Ref.~\cite{Cohen:1998zx}, inspired by the Bekenstein-Hawking entropy bound in black hole thermodynamics \cite{Bekenstein:1973ur,Bekenstein:1974ax,Hawking:1975vcx,Hawking:1976de,Bekenstein:1980jp,Bekenstein:1993dz}, postulate that the energy within a region of size $L$ should not exceed the mass of a black hole of the same size, thus $L^3\rho\le LM_P^{2}$. In a cosmological context, the largest allowable scale for $L$ saturates this inequality. Since these seminal works, several dark energy models rooted in the holographic principle have been explored. For example, it has been demonstrated that selecting the scale $L$ as the Hubble length \cite{Hsu:2004ri} or the size of the particle horizon \cite{Li:2004rb} does not lead to accelerated expansion. A successful model where the scale $L$ is determined by the size of the future event horizon was proposed in Ref. \cite{Li:2004rb}, however, this model has faced criticism due to causality issues \cite{Cai:2007us}.

In Ref.~\cite{Gao:2007ep} the holographic Ricci dark energy model (HRDE) was proposed, a model that circumvents the causality problems, where the dark energy density is proportional to the Ricci scalar. Subsequently, in Ref. \cite{Cai:2008nk}, it was recognized that the Jeans length of perturbations establishes a causal connection scale dictated by the Ricci scalar, offering a physical motivation for the HRDE model. Numerous studies have delved into these models, as evidenced by references such as \cite{Xu:2008rp,Feng:2008kz,Feng:2008rs,Li:2009bn,Zhang:2009un,Lepe:2010vh,Kim:2010pdn,delCampo:2011jp,delCampo:2013hka}.

In particular, the authors of Ref. \cite{Granda:2008dk} propose a new scale (infrared cut-off) for the holographic dark energy model which includes a term proportional to $\dot{H}$, 
\begin{equation}
\rho_x= 3(\alpha H^2+\beta \dot{H})\label{EOTH}  
\end{equation}
where $\alpha$ and $\beta$ are constants. This model, dubbed the modified holographic Ricci dark energy model (MHRDE) has been widely studied (see for example Refs.\cite{Granda:2009xu,Granda:2009di,Wang:2010kwa,Mathew:2012md}). In particular, the authors of Ref. \cite{bib20} indicate that the scale $L$ leading to \eqref{EOTH} is a natural extension of the HDE model. Notice that for $\alpha=2\beta$, the HRDE model is recovered for a flat scenario from \eqref{EOTH}. 

Interacting holographic dark energy models have been extensively studied, originally with the aim of find an accelerating scenario with $L=H$ \cite{Pavon:2005yx}. Over the years, many interacting holographic scenarios have been studied in different contexts, see for example references \cite{Hu:2006ar,Fu:2011ab,Chimento:2011dw,Chimento:2011pk,Chimento:2012zz,Chimento:2012hn,Chimento:2013se,P:2013cmq,Arevalo:2013tta,Chimento:2013qja,Oliveros:2014kla,Som:2014hja,Mahata:2015nga,Pan:2014afa,Arevalo:2014zoa,Lepe:2015qhq,Zadeh:2016vgc,Herrera:2016uci,Feng:2016djj,George:2019vko}.

Bayesian model selection is a powerful statistical tool in comparing the performance of cosmological models, in light of the available data. It has been widely used in cosmology \cite{Santos:2016sti,Heavens:2017hkr,SantosdaCosta:2017ctv,Andrade:2017iam,Ferreira:2017yby,Cid:2018ugy}. In particular, in Ref.~\cite{Cid:2020kpp} it was performed a Bayesian model selection for MHRD-IDE models with linear interaction. This work can
be considered as an extension of Ref.~\cite{Cid:2020kpp}, with five new types of interactions revised.

This paper investigates the viability of a class of interacting scenarios in the framework of modified holographic Ricci dark energy models. To assess the models' viability we characterize the asymptotic behavior of MHR-IDE models through a dynamical system analysis and study its performance in fitting the known asymptotic behavior of our universe. Furthermore, we conduct Bayesian model selection, comparing the performance of MHR-IDE scenarios with the $\Lambda$CDM model, using background data such as supernovae type Ia, cosmic chronometers, the local value of the Hubble constant, baryon acoustic oscillations and the angular scale of the sound horizon at the last scattering. 

The paper's structure is as follows.
In section \ref{sec2} we present the studied MHR-IDE scenarios. In section \ref{sec3} we develop the dynamical system analysis, including critical points, existence and stability conditions, and a description of each scenario. In \ref{sec4} we describe the data used and the methodology to assess the Bayesian comparison. In section \ref{sec5} we discuss the main results of this work and finally, in section \ref{remarks} we present the final remarks on this work.

\section{The modified holographic Ricci interacting dark energy models}\label{sec2}

We work in the framework of General Relativity by considering a spatially flat Friedmann-Lemaître-Robertson-Walker (FLRW) metric. We assume a universe composed of baryons, radiation, dark matter, and dark energy, where each of these components is considered as barotropic perfect fluids. In this scenario, the Friedmann equations are written as,
\begin{align}\label{eq1}
    3H^2 = \rho \\
    2\dot{H} + 3H^2 = -p
\end{align}
where $\rho = \rho_b + \rho_r + \rho_c + \rho_x$ and $p = p_b + p_r + p_c + p_x$ are the total energy density and pressure, respectively, with the subscripts $b$, $r$, $c$, $x$ for baryons, radiation, cold dark matter and dark energy. $H=\dot{a}/a$ is the Hubble expansion rate defined in terms of the scale factor $a$, the dot represents a derivative with respect to the cosmic time and we use units such as $8\pi G = 1$ and $c=1$. On the other hand, from the energy-momentum tensor conservation, we have
\begin{equation}\label{eq3}
    \dot{\rho} + 3H(\rho+p)  = 0. 
\end{equation}

Although the total energy density is conserved, this does not imply that each component is conserved separately, which allows us to assume that the dark components interact with each other through a phenomenological interaction term $Q$. Thus, considering a barotropic equation of state for all the components, $p_i = \omega_i\rho_i$, where $\omega_i$ is the state parameter, we split Eq. \eqref{eq3} into the following equations.
\begin{align}\label{eq4}
    \dot{\rho}_b + 3H\rho_b &= 0 \\ \label{eq5}
    \dot{\rho}_r + 4H\rho_r &= 0 \\ \label{eq6}
    \dot{\rho}_c + 3H\rho_c &= -Q \\ \label{eq7}
    \dot{\rho}_x + 3H(1+\omega_x)\rho_x &= Q 
\end{align}
where we assume $\omega_b = \omega_c = 0, \omega_r = 1/3$ and $\omega_x$ is a variable dark energy state parameter. Using  the change of variable $\eta = 3\ln a$ and defining $()' = d/d\eta$, Eqs.~\eqref{eq4}-\eqref{eq7} are rewritten as
\begin{align} \label{eq8}
    \rho'_b + \rho_b &= 0 \\ \label{eq9}
    \rho'_r + \frac{4}{3}\rho_r &= 0 \\ \label{eq10}
    \rho'_c + \rho_c &= -\Gamma \\ \label{eq11}
    \rho'_x + (1+\omega_x)\rho_x &= \Gamma
\end{align}
with $\Gamma = Q/3H$. Note that for $\Gamma>0$, we have an energy transfer from cold dark matter to dark energy and for $\Gamma<0$ we have the opposite energy transfer.

Taking into account the holographic motivation for the dark energy discussed in section \ref{sec1}, we consider the general form \eqref{EOTH} for the dark energy energy density written as,
\begin{equation} \label{eq12}
\rho_x = \alpha \rho + \frac{3\beta}{2}\rho',
\end{equation}
where $\alpha$ and $\beta$ are considered as positive constants. For this scenario, the authors of Ref. \cite{bib15} get the following second order differential equation by combining Eqs.~\eqref{eq10}-\eqref{eq12},
\begin{align}\label{eq13}
    \frac{3\beta}{2}\rho''_d + \left(\alpha + \frac{3\beta}{2} - 1\right)\rho'_d + (\alpha-1)\rho_d + \frac{1}{3}(2\beta - \alpha)\rho_{r0}e^{-4\eta/3} = \Gamma
\end{align}
where $\rho_d = \rho_c + \rho_x$ and $\rho_{r0}$ is the integration constant from Eq.(\ref{eq9}).

In this work, we study five types of modified holographic Ricci dark energy models (MHRDE), defined as: 
\begin{eqnarray*}
\Gamma_1 = \delta \rho'_c + \gamma \rho'_x,\quad
\Gamma_2 = \delta \rho_d + \gamma \rho'_d,\\
\Gamma_3 = \gamma \rho,\quad\;\,
\Gamma_4 = \gamma \rho',\quad\;\,
\Gamma_5 = \gamma q \rho, 
\end{eqnarray*}
where $q = -(1+\dot{H}/H^2)$ is the deceleration parameter, $\rho$ is the total energy density and $\delta$, $\gamma$ are constants. We can rewrite Eq.(\ref{eq13}) as
\begin{align}\label{eq14}
\rho''_d + b_1 \rho'_d + b_2 \rho_d + b_3 \Omega_{b0}a^{-3} + b_4 \Omega_{r0}a^{-4} = 0.
\end{align}

This equation includes all the proposed interactions, so that $b_1$, $b_2$, $b_3$, and $b_4$ are different combinations of the holographic and interaction parameters (see Table \ref{tab1}). 

\begin{table}[htb]
\centering
\begin{tabular}{ c  c  c  c  c }
\hline
Model & $b_1$ & $b_2$ & $b_3$ & $b_4$  \\ \hline
 &  &  &  &  \\
$\Gamma_1 $ & $\displaystyle\frac{2\alpha(1+\delta-\gamma)-2-2\delta+3\beta}{3\beta(1+\delta-\gamma)}$ & $\displaystyle\frac{2(\alpha-1)}{3\beta(1+\delta-\gamma)}$ & $\displaystyle\frac{(2\alpha-3\beta)(\gamma-\delta)}{3\beta(1+\delta-\gamma)}$ & $-\displaystyle\frac{2(\alpha-2\beta)(1+4\delta-4\gamma)}{9\beta(1+\delta-\gamma)}$  \\
 &  &  &  &  \\
$\Gamma_2 $ & $\displaystyle\frac{2\alpha+3\beta-2(\gamma+1)}{3\beta}$ & $\displaystyle\frac{2}{3\beta}(\alpha-\delta-1)$ & $0$ & $-\displaystyle\frac{2}{9\beta}(\alpha-2\beta)$  \\
 &  &  &  &  \\
$\Gamma_3$ & $1 + \displaystyle\frac{2(\alpha-1)}{3\beta}$ & $\displaystyle\frac{2}{3\beta}\displaystyle(\alpha-\gamma-1)$ & $-\displaystyle\frac{2\gamma}{3\beta}$ & $-\displaystyle\frac{2}{9\beta}\displaystyle(\alpha - 2\beta+ 3\gamma )$ \\
 &  &  &  &  \\
$\Gamma_4$ & $1 + \displaystyle\frac{2(\alpha-\gamma - 1)}{3\beta}$ & $\displaystyle\frac{2}{3\beta}\displaystyle(\alpha-1)$ & $\displaystyle\frac{2\gamma}{3\beta}$ & $-\displaystyle\frac{2}{9\beta}\displaystyle(\alpha-2\beta-4\gamma)$ \\
 &  &  &  &  \\
$\Gamma_5$ & $1+\displaystyle\frac{\gamma}{\beta} + \displaystyle\frac{2}{3\beta}(\alpha-1)$ & $\displaystyle\frac{2}{3\beta}(\alpha + \gamma-1)$ & $-\displaystyle\frac{\gamma}{3\beta}$ & $\displaystyle-\frac{2}{9\beta}(\alpha- 2\beta+3\gamma )$ \\
 &  &  &  &  \\ \hline
\end{tabular}

\caption{Definition of the constants $b_1$, $b_2$, $b_3$ and $b_4$ in terms of the parameters defined for each MHR-IDE scenario.}\label{tab1}
\end{table}

The general solution of the Eq. \eqref{eq14} takes the general form
\begin{equation}\label{eq15}
    \rho_d(a) = 3H_0^2( \Omega_{h1}a^{-4} +\Omega_{h2}a^{-3} + \tilde{C}_1a^{3\lambda_1} + \tilde{C}_2a^{3\lambda_2}),
\end{equation}
where the coefficients and integration constants in \eqref{eq15} are given by
\begin{align}
\Omega_{h1} &= \frac{9b_4\Omega_{r0}}{12b_1 - 9b_2 - 16},\quad\Omega_{h2} = \frac{b_3\Omega_{b0}}{b_1 - b_2 - 1}\label{eq18}\\
\tilde{C}_1 &= \frac{\Omega_{h2}(1+\lambda_2)}{\lambda_1 - \lambda_2} + \frac{\Omega_{h1}(4+3\lambda_2)}{3(\lambda_1 - \lambda_2)} + \frac{2(\Omega_{x0}-\alpha)}{3\beta(\lambda_1 - \lambda_2)} + \frac{3\Omega_{b0} + 4\Omega_{r0} - 3\lambda_2(\Omega_{c0} + \Omega_{x0})}{3(\lambda_1 - \lambda_2)}\label{eq16}\\ 
\tilde{C}_2 &= -\Omega_{h1}-\Omega_{h2}+\Omega_{c0} + \Omega_{x0}-\tilde{C}_1\label{eq17b}
\end{align}
and
\begin{equation}\label{eq17}
    \lambda_{1,2} = -\frac{1}{2}(b_1 \pm \sqrt{{b_1}^2 - 4b_2})
\end{equation}
where $\Omega_{c0},\Omega_{x0}$ are the current values of the density parameters for cold dark matter and the MHRDE, respectively.

Therefore, we can write the Hubble expansion rate associated to the total energy density as:
\begin{equation}\label{eq19}
H(a) = H_0\sqrt{(\Omega_{h1} + \Omega_{r0})a^{-4} + (\Omega_{h2}+\Omega_{b0})a^{-3} + \tilde{C}_1a^{3\lambda_1} + \tilde{C}_2a^{3\lambda_2}}.
\end{equation}


\section{Dynamical System Analysis}\label{sec3}
We apply dynamical system methods \cite{Copeland:2006wr,Bahamonde:2017ize} to identify the relevant cosmological eras in the MHR-IDE models.
By writing the system of Eqs. \eqref{eq8}-\eqref{eq11} in terms of the density parameters $\Omega_i=\rho_i/3H^2$ and using the Friedmann equation \eqref{eq1} as a constraint among density parameters, i.e.,
\begin{equation}
\Omega_r+\Omega_b+\Omega_c+\Omega_x=1.\label{const0}
\end{equation}
With this constraint we reduce  the system of equations \eqref{eq8}-\eqref{eq11} to three equations for,  $\Omega_r'$, $\Omega_c'$ and $\Omega_x'$, where $\Omega_r'$ is given by
\begin{equation}
\Omega_r'=\frac{2 \Omega_r (\alpha -2 \beta -\Omega_x)}{3 \beta },\label{rhop}
\end{equation}
and it has the same structure for all the MHR-IDE models. The remaining equations, $\Omega_c'$  and $\Omega_x'$ depend upon the $\Gamma$ model and are shown for each model in the following subsections.

The stability of the critical points representing the different cosmological epochs is found by calculating the eigenvalues of the linearized system at the critical points. For each studied scenario, we consider the set of equations,
\begin{eqnarray}
\Omega_i'=f_i(\Omega_l),\label{DS}
\end{eqnarray}
where $f_i$ is a function of the density parameters $\Omega_r$, $\Omega_c$ and $\Omega_x$ and $i=r,c,x$. From Eq. \eqref{DS} we find the critical points $\Omega_l^*$ by calculating 
\begin{equation*}
f_i(\Omega_l^*)=0.
\end{equation*}
Then we linearized the set of equations \eqref{DS} around the critical points,
\begin{eqnarray*}
\delta \Omega_i'=J_i^l(\Omega_j^*)\Omega_l,
\end{eqnarray*}
where $J_i^l=\frac{\partial f_i}{\partial \Omega_l}$ is the Jacobian matrix, from which we can find different regions in the parameter space. In particular, we have unstable, saddle, or stable points when the real part of the eigenvalues are, all positives, a mixture of positives and negatives, or all negatives, respectively.
\vspace{0.3cm}

We perform a dynamical system analysis of models $\Gamma_1-\Gamma_5$ presented in Table \ref{tab1}. We study models $\Gamma_1$ and $\Gamma_2$ in four possible scenarios, models $\Gamma_{11}$, $\Gamma_{12}$, $\Gamma_{13}$ and $\Gamma_{14}$ representing particular cases of $\Gamma_1$ for $\delta=0$, $\gamma=0$, $\delta = \gamma$ and $\delta \neq \gamma$, respectively. Similarly, $\Gamma_{21}$, $\Gamma_{22}$, $\Gamma_{23}$ and $\Gamma_{24}$ are the cases of $\Gamma_2$ for the same scenarios. In Tables \ref{T2}--\ref{T4} we show the critical points and the corresponding stability conditions and classification resulting from the dynamical system analysis for scenarios $\Gamma_1$, $\Gamma_2$ and $\Gamma_3-\Gamma_5$, respectively. The scenarios $\Gamma_{14}$ and $\Gamma_{24}$ are too intricate to get significant results, so we limit to cases $1-3$ in models $\Gamma_1$ and $\Gamma_2$. To improve the visualization of the critical points we have defined constants $C_1-C_7$ in Table \ref{T2},
\begin{eqnarray}
C_1=\alpha-2\beta, \quad
C_{(2,3)}=1+\alpha\gamma-C_4\pm\sqrt{(1+\alpha\gamma-C_4)^2-6\beta\gamma}, \nonumber \\
C_4=\frac{1}{2}(2\alpha-3\beta), \quad
C_5=1+2\beta+4\gamma, \quad
C_{(6,7)}=-\gamma-C_4\pm\sqrt{(1+\gamma-C_4)^2-6\beta\gamma}.\label{Const1}
\end{eqnarray}
Constants $D_1-D_7$ in Table \ref{T3},
\begin{eqnarray}
D_{(1,2)}=C_4\mp\sqrt{(1-C_4)^2+6\beta\delta},\quad D_3=-\frac{(1-C_4)^2}{6\beta},\quad D_4=\frac{1}{2}(5-2\sqrt{6}),\nonumber\\
D_5=\frac{1}{2}(2+2\gamma+3\beta),\quad D_{(6,7)}=\gamma-C_4\pm\vert1+\gamma-C_4\vert,\label{Const2}
\end{eqnarray}
and constants $E_1-E_3$ in Table \ref{T3},
\begin{eqnarray}
E_1=-\frac{1}{4}(1+2\beta),\quad E_{(2,3)}=C_4\mp\sqrt{(C_4-1+\frac{3}{2}\gamma)^2+3\beta\gamma}. \label{Const3}
\end{eqnarray}

On the other hand, in order to analyze the type of the effective fluid for each critical point, in Tables \ref{T2}--\ref{T4} we use the effective state parameter defined as $\omega_{\textrm{eff}}=\frac{p}{\rho}$. In the following, we describe the critical points and the existence and stability conditions for each model. Also, to develop the analysis of the stability conditions, we restrict ourselves to the following parameter ranges 
\begin{equation}
0<\alpha<1,\quad 0<\beta<1,\quad -1<\gamma<1,\quad -1<\delta<1, \label{ranges}
\end{equation}
which makes sense in the context of MHR-IDE models.

\begin{table*}[!htb]
\begin{center}
\begin{minipage}{\textwidth}
\begin{tabular}{ccccc}
&Critical Points & $\omega_{\rm eff}$ &  Stability Conditions&Classification \\\hline
\multicolumn{5}{c}{$ \displaystyle\Gamma_{11}$}\\\hline
\\[-8pt]
\multirow{6}{*}{$P_1$}&
\multirow{6}{*}{$\{1-C_1(1-4\gamma),-4\gamma C_1,C_1\}$} & 
\multirow{6}{*}{$\displaystyle\frac{1}{3}$} &$\alpha>\frac{1}{5}$, $-\frac{1}{3}\le C_1\le \frac{1}{5}$& unstable\\
&&&$\alpha\le\frac{1}{5}$, $C_1\ge-\frac{1}{3}$  &unstable\\
&&&$C_1(1-4\gamma)<1$, $C_1<-\frac{1}{3}$        &unstable\\
&&&$C_1(1-4\gamma)<1$, $C_1>\frac{1}{5}$         &unstable\\
&&&$C_1(1-4\gamma)>1$, $C_1<-\frac{1}{3}$        &saddle\\
&&&$C_1(1-4\gamma)>1$, $C_1>\frac{1}{5}$         &saddle\\[2pt]
\cline{1-5}
\\[-6pt]

\multirow{9}{*}{$P_2$}&
\multirow{9}{*}{$\displaystyle\left\{0,\frac{C_2-2\gamma}{2(1-\gamma)},\frac{2-C_2}{2(1-\gamma)}\right\}$} & 
\multirow{9}{*}{$\displaystyle\frac{3\beta\gamma-C_3}{3\beta(1-\gamma)}$}& $C_1(1-4\gamma)>1$, $C_1<-\frac{1}{3}$                                                            &unstable\\
&&&$C_1(1-4\gamma)>1$, $C_1>\frac{1}{5}$   &unstable\\
&&& $\gamma>0$, $2\alpha>3\beta$           &saddle\\
&&&$\gamma<0$, $2\alpha<3\beta$            &saddle\\
&&&$\gamma<0$, $\alpha\le\frac{1}{5}$, $2\alpha>3\beta$                                                       &saddle\\
&&&$\gamma<0$, $\alpha>\frac{1}{5}$, $C_1\le\frac{1}{5}$, $2\alpha>3\beta$  &saddle\\
&&&$\gamma<0$, $C_1(1-4\gamma)<1$, $C_1>\frac{1}{5}$                                 &saddle\\
&&&$\gamma>0$, $C_1\ge-\frac{1}{3}$, $2\alpha<3\beta$                               & saddle\\
&&&$\gamma>0$, $C_1(1-4\gamma)<1$, $C_1<-\frac{1}{3}$                               & saddle\\[1pt] \hline
\\[-8pt]

$P_3$&$\displaystyle\left \{0,\frac{C_3-2\gamma}{2(1-\gamma)},\frac{2-C_3}{2(1-\gamma)}\right\}$ &$\displaystyle\frac{3\beta\gamma-C_2}{3\beta(1-\gamma)}$& -& stable\\[8pt] \hline\cline{1-5}

\\[-6pt]
\multicolumn{5}{c}{$ \displaystyle\Gamma_{12}$}\\\hline
\multirow{2}{*}{$P_4$}&
\multirow{2}{*}{$\displaystyle\left\{1-C_1,0,C_1\right\}$} & 
\multirow{2}{*}{$\displaystyle\frac{1}{3}$} & $\delta>-\frac{1}{4}$& unstable \\[3pt]
&&&$ \delta<-\frac{1}{4}$&saddle\\[3pt]\hline
\\[-8pt]

$P_5$&$\displaystyle\left\{0,0,C_4\right\}$ & $\displaystyle0$&$ \delta<0$ or $\delta>0$ & saddle \\[1pt]\hline

\multirow{2}{*}{$P_6$}&
\multirow{2}{*}{$\displaystyle\left\{0,1-\frac{C_4+\alpha\delta}{(1+\delta)},\frac{C_4+\alpha\delta}{(1+\delta)}\right\}$} & 
\multirow{2}{*}{$\displaystyle-\frac{\delta}{1+\delta}$} & $\delta<-\frac{1}{4}$& unstable\\
&&&$ -\frac{1}{4}<\delta<0$ or  $\delta>0$&saddle\\[3pt] \hline

$P_7$&$\displaystyle\{0,0,1\}$ & $\displaystyle\frac{2C_4-2}{3\beta}$& -&stable\\ \hline\cline{1-5}
\\[-6pt]

\multicolumn{5}{c}{$ \displaystyle\Gamma_{13}$}\\\hline
\multirow{4}{*}{$P_8$}&
\multirow{4}{*}{$\displaystyle\left\{\frac{1+4\gamma-C_1}{1+4\gamma},-\frac{4\gamma C_1}{1+4\gamma},C_1\right\}$} & 
\multirow{4}{*}{$\displaystyle\frac{1}{3}$} & $\beta<\frac{2}{3}$, $\alpha<C_5\le1$& unstable\\
&&&$\beta<\frac{2}{3}$, $C_5>1$&unstable \\
&&&$\beta\ge\frac{2}{3}$,  $\alpha<C_5<1$ or $\beta\ge\frac{2}{3}$, $C_5\ge1$&unstable \\
&&&$C_5\le0$ or $0<C_5<\alpha$&saddle  \\[3pt]
\cline{1-5}
\\[-6pt]
$P_9$ &$\displaystyle\left\{0,-C_4, C_4\right\}$ & $\displaystyle0$& $ \gamma<0$ or $\gamma>0$& saddle\\[2pt]
\cline{1-5}
\\[-6pt] 

\multirow{6}{*}{$P_{10}$}&
\multirow{6}{*}{$\displaystyle\left\{0,\frac{1}{2}(1+C_6),\frac{1}{2}(1-C_6)\right\}$} & 
\multirow{6}{*}{$\displaystyle-\frac{1+C_7+2\gamma}{3\beta}$} & $C_5\le0$ or $0<C_5<\alpha$& unstable \\
&&&$ \gamma>0$&saddle\\
&&&$ \beta<\frac{2}{3}$, $\alpha<C_5\le1$&saddle \\
&&&$ \beta<\frac{2}{3}$, $-\frac{\beta}{2}<\gamma<0$ &saddle\\
&&&$\beta\ge \frac{2}{3}$, $\alpha<C_5<1$ &saddle\\
&&&$\beta\ge \frac{2}{3}$, $-\frac{\beta}{2}<\gamma<0$&saddle  \\[3pt] 
\cline{1-5}
\\[-6pt] 

$P_{11}$ & $\displaystyle\left\{0,\frac{1}{2}(1+C_7),\frac{1}{2}(1-C_7)\right\}$ & $\displaystyle-\frac{1+C_6+2\gamma}{3\beta}$& -&stable\\[6pt]\hline
\end{tabular}
\end{minipage}
\end{center}
\caption{\label{T2} Description of the critical points $(\Omega_{r}^{*},\Omega_{c}^{*},\Omega_{x}^{*})$ for scenario $\Gamma_{1}$. The ranges in \eqref{ranges} are considered as global constraints. To improve visualization we use the constants $C_1-C_7$ defined in \eqref{Const1}.}
\end{table*}
\subsection{Scenarios $\Gamma_1$}
For the model $\Gamma_1$ we have, besides Eq.~\eqref{rhop}, the following set of equations,
\begin{eqnarray}
\Omega_c'&=&\frac{2 (\alpha-\Omega_x) ((\delta+1-\gamma) \Omega_c+\gamma)+\beta  (\gamma (3 \Omega_c-\Omega_r+3 \Omega_x-3)-3 \Omega_c)}{3 \beta  (\delta-\gamma+1)},\\
\Omega_x'&=&\frac{2 (\alpha-\Omega_x)  ((\delta+1-\gamma)  (\Omega_x-1)-\gamma )+\beta  (-\delta  (3 \Omega_c-\Omega_r+3 \Omega_x-3)
-3 \Omega_x+\Omega_r+3)
}{3 \beta  (\delta -\gamma+1)}.
\end{eqnarray}
Inside the range \eqref{ranges}, the constants $C_1$, $C_2$, $C_6$ and $C_7$ in Table \ref{T2} are real numbers and we must have $\gamma\neq -\frac{1}{4}$ in order to the critical point $P_8$ exists. In the following we describe the critical points $P_1-P_{11}$ in Table \ref{T2}, corresponding to models $\Gamma_{11}-\Gamma_{13}$.
\vspace{0.2cm}

For the model $\Gamma_{11}$ we find three critical points, $P_1-P_3$. The point $P_1$ is a combination of radiation and the dark components and corresponds to an effective fluid of radiation type. Positive energy densities at this critical point require $C_1>0$ and $\gamma<0$. Notice that for $C_1\ll1$ we get the dominance of the radiation term at $P_1$. 
The points $P_2$ and $P_3$ correspond to a combination of the dark components. At the point $P_3$ the dark energy term dominates and $\omega_{\textrm{eff}}<-1$ in the range \eqref{ranges}. For $\gamma\ll1$ we have $\omega_{\textrm{eff}}\approx \frac{C_4 \gamma}{C_4-1}$ at $P_2$ and $\Omega_x^*\approx1$ at $P_3$. 
\vspace{0.2cm}

For the model $\Gamma_{12}$ we find four critical points, $P_4-P_7$ in Table \ref{T2}. The point $P_4$ is an effective fluid of radiation type, corresponding to a combination of the radiation component with the dark energy component (DE). $P_5$ is a saddle point corresponding to a combination of baryons and DE with $\omega_{\textrm{eff}}=0$, for $C_4\approx0$ this point corresponds to the dominance of baryons. $P_6$ is a combination of the dark components, with $\omega_{\textrm{eff}}\approx0$ if $\delta\ll1$. $P_7$ represents the dominance of DE with $\omega_{\textrm{eff}}<-1$ in the range \eqref{ranges}.
\vspace{0.2cm}

For the model $\Gamma_{13}$ we find four critical points, $P_8-P_{11}$ in Table \ref{T2}. The point $P_8$ is a combination of radiation and the dark components and it corresponds to an effective fluid of radiation type. Positive energy densities at the critical points require $C_1>0$ and $\gamma<0$. $P_9$ represents a combination of the dark sector and baryons and, at this critical point, we always have one negative energy density and $\omega_{\textrm{eff}}=0$. For $C_4\approx0$ this point corresponds to the dominance of baryons. The points $P_{10}$ and $P_{11}$ are combinations of the dark sector. The point $P_{11}$ has $\omega_{\textrm{eff}}<-1$ in the range \eqref{ranges}. For $\gamma\ll1$ we have $\omega_{\textrm{eff}}\approx \frac{\gamma}{C_4-1}$ at $P_{10}$ and $\Omega_x^*\approx1$ at $P_{11}$.
 
\begin{table*}[!htb]
\begin{center}\begin{minipage}{\textwidth}
\begin{tabular}{ccccc}

&Critical Points & $\omega_{\rm eff}$ & Stability Conditions & Classification\\\hline
\multicolumn{5}{c}{$ \Gamma_{21}$}\\\hline
\\[-8pt]

\multirow{4}{*}{$Q_1$}&
\multirow{4}{*}{$\displaystyle\left\{1-\frac{C_1}{1+4\gamma},-\frac{4\gamma C_1}{1+4\gamma},C_1\right\}$} & 
\multirow{4}{*}{$\displaystyle\frac{1}{3}$}&
$\beta<\frac{2}{3}$, $\alpha<C_5\le1$ &unstable\\
&&& $\beta<\frac{2}{3}$, $\beta>-2\gamma$&unstable\\
&&& $\beta\ge\frac{2}{3}$, $\alpha<C_5<1$ or $\beta\ge\frac{2}{3}$, $\beta>-2\gamma$&unstable\\
&&& $C_5\le0$ or $\alpha<C_5<1$&saddle\\[3pt]
\cline{1-5}
\\[-8pt]

$Q_2$&$\displaystyle\{0,-C_4,C_4\}$ & $\displaystyle0$ & $\gamma>0$ or $\gamma<0$&unstable \\[2pt]
\cline{1-5}
\\[-6pt]

\multirow{4}{*}{$Q_3$}&
\multirow{4}{*}{$\displaystyle\left\{0,\frac{1}{2}(1+C_6),\frac{1}{2}(1-C_6)\right\}$} &
\multirow{4}{*}{$\displaystyle-\frac{1+C_7+2\gamma}{3\beta}$}& $C_5\le0$ or $0<C_5\le1$&unstable\\
&&& $\beta<\frac{2}{3}$, $\alpha<C_5\le1$ or $\gamma>0$&saddle\\
&&& $\beta<\frac{2}{3}$, $-\frac{\beta}{2}<\gamma<0$&saddle\\
&&& $\beta\ge\frac{2}{3}$, $\alpha<C_5\le1$ or $\beta\ge\frac{2}{3}$, $-\frac{\beta}{2}<\gamma<0$&saddle\\[3pt]
\cline{1-5}
\\[-6pt]

$Q_4$&$\displaystyle\left\{0,\frac{1}{2}(1+C_7),\frac{1}{2}(1-C_7)\right\}$ & $\displaystyle-\frac{1+C_6+2\gamma}{3\beta}$&-& stable\\\hline\cline{1-5}
\\[-6pt]

\multicolumn{5}{c}{$ \Gamma_{22}$}\\\hline
\multirow{3}{*}{$Q_5$}&
\multirow{3}{*}{$\displaystyle\left\{\frac{1-3\delta-C_1}{1-3\delta},\frac{3\delta C_1}{1-3\delta},C_1\right\}$} & 
\multirow{3}{*}{$\displaystyle\frac{1}{3}$} &$1-C_4\ge\sqrt{6\beta}$, $\delta<\frac{1}{3}(1-C_1)$& unstable \\
&&& $1-C_4<\sqrt{6\beta}$, $D_3\le\delta<\frac{1}{3}(1-C_1)$&unstable\\ 
&&& $\delta>\frac{1}{3}(1-C_1)$&saddle\\[3pt] \hline

\multirow{2}{*}{$Q_6$}&
\multirow{2}{*}{$\displaystyle\left\{0,-C_4,C_4\right\}$} & 
\multirow{2}{*}{$\displaystyle0$}&$1-C_4\ge\sqrt{6\beta}$, $\delta<0$ or $\delta>0$&saddle\\
&&& $1-C_4<\sqrt{6\beta}$, $D_3\le\delta<0$&saddle\\[3pt]
\cline{1-5}
\\[-6pt]

\multirow{4}{*}{$Q_7$}&
\multirow{4}{*}{$\displaystyle\left\{0,\frac{1}{2}(1-D_1),\frac{1}{2}(1+D_1)\right\}$} & 
\multirow{4}{*}{$\displaystyle\frac{D_2-1}{3\beta}$} & $\delta>\frac{1}{3}(1-C_1)$&  unstable\\
&&& $1-C_4\ge\sqrt{6\beta}$, $\delta<0$&saddle\\
&&& $1-C_4<\sqrt{6\beta}$, $D_3<\delta<0$&saddle\\
&&& $0<\delta<\frac{1}{3}(1-C_1)$&saddle\\[3pt]
\cline{1-5}
\\[-6pt]

\multirow{5}{*}{$Q_8$}&
\multirow{5}{*}{$\displaystyle\left\{0,\frac{1}{2}(1-D_2),\frac{1}{2}(1+D_2)\right\}$} & 
\multirow{5}{*}{$\displaystyle\frac{D_1-1}{3\beta}$}& $\alpha<D_4$, $1+C_4\ge2\sqrt{\alpha}$& stable\\
&&& $\alpha<D_4$, $1+C_4\le-2\sqrt{\alpha}$&stable \\
&&& $\alpha<D_4$, $-2\sqrt{\alpha}<1+C_4<2\sqrt{\alpha}$, $\delta>D_3$&stable\\
&&& $\alpha\ge D_4$, $1+C_4\ge2\sqrt{\alpha}$&stable\\
&&& $\alpha\ge D_4$, $1+C_4<  2\sqrt{\alpha}$, $\delta>D_3$&stable\\\hline
\cline{1-5}
\\[-6pt]

\multicolumn{5}{c}{$ \Gamma_{23}$}\\\hline
\multirow{4}{*}{$Q_9$}&
\multirow{4}{*}{$\displaystyle\left\{\frac{1+\gamma-C_1}{1+\gamma},-\frac{\gamma C_1}{1+\gamma},C_1\right\}$} & 
\multirow{4}{*}{$\displaystyle\frac{1}{3}$} & $\beta\le\frac{1}{2}$, $\gamma\le-2\beta$, $C_1<1+\gamma$& unstable\\
&&& $\beta\le\frac{1}{2}$, $\gamma>-2\beta$ &unstable\\
&&& $\beta>\frac{1}{2}$&unstable\\
&&& $\beta<\frac{1}{2}$, $\gamma<-2\beta$, $C_1>1+\gamma$&saddle\\[3pt]
\cline{1-5}
\\[-6pt]

$Q_{10}$&$\displaystyle\left\{0,-C_4,C_4\right\}$ & $\displaystyle0$&-& non-hyperbolic \\\hline

\multirow{4}{*}{$Q_{11}$}&
\multirow{4}{*}{$\displaystyle\left\{0,\frac{1}{2}(1-D_7),\frac{1}{2}(1+D_7)\right\}$} & 
\multirow{4}{*}{$\displaystyle-\frac{1-D_6+2\gamma}{3\beta}$} & $\beta<\frac{1}{2}$, $\gamma<-2\beta$, $1+\gamma<C_1$&unstable\\
&&& $\beta\le\frac{1}{2}$, $\gamma\le-2\beta$, $D_5<\alpha<1+\gamma+2\beta$, &saddle\\
&&& $\beta\le\frac{1}{2}$, $-2\beta<\gamma<-\frac{3\beta}{2}$, $\alpha>D_5$&saddle\\
&&& $\frac{1}{2}<\beta<\frac{2}{3}$, $\gamma<-\frac{3\beta}{2}$, $\alpha>D_5$&saddle\\[3pt] \hline

\multirow{4}{*}{$Q_{12}$}&
\multirow{4}{*}{$\displaystyle\left\{0,\frac{1}{2}(1-D_6),\frac{1}{2}(1+D_6)\right\}$ }&
\multirow{4}{*}{$\displaystyle-\frac{1-D_7+2\gamma}{3\beta}$}& $\beta<\frac{1}{2}$, $\gamma<-\frac{3\beta}{2}$, $\alpha<D_5$& stable\\
&&& $\beta<\frac{1}{2}$, $\gamma>-\frac{3\beta}{2}$&stable\\
&&& $\frac{1}{2}\le\beta<\frac{2}{3}$, $\gamma<-\frac{3\beta}{2}$, $\alpha<D_5$&stable\\
&&& $\frac{1}{2}\le\beta<\frac{2}{3}$, $\gamma\ge-\frac{3\beta}{2}$ or $\beta>\frac{2}{3}$&stable\\[3pt] \hline

\end{tabular}
\end{minipage}
\end{center}
\caption{\label{T3} Description of the critical points $(\Omega_{r}^*,\Omega_{c}^*,\Omega_{x}^*)$ for scenario $\Gamma_{2}$. The ranges in \eqref{ranges} are considered as global constraints. To improve visualization we use the constants $D_1-D_7$ defined in \eqref{Const2}.}

\end{table*}

\subsection{Scenarios $\Gamma_2$}
For the model $\Gamma_2$ we have, besides Eq.~\eqref{rhop}, the following set of equations,
\begin{eqnarray}
\Omega_c'&=&\frac{2 \alpha  (\gamma+\Omega_c)+\beta  (\gamma (3 \Omega_c-\Omega_r+3 \Omega_x-3)-3 (\delta \Omega_c+\delta \Omega_x+\Omega_c))-2 \Omega_x (\gamma+\Omega_c)}{3 \beta },\\
\Omega_x'&=&\frac{2 \alpha  (\Omega_x-1-\gamma)+\beta  (3 (\delta-\gamma) (\Omega_c+\Omega_x)+(\gamma+1) (\Omega_r+3)-3 \Omega_x)+2 \Omega_x (\gamma-\Omega_x+1)}{3 \beta }.
\end{eqnarray}

In the following we describe critical points $Q_1-Q_{12}$ in Table \ref{T3}, corresponding to models $\Gamma_{21}-\Gamma_{23}$.
\vspace{0.2cm}

In the range \eqref{ranges}, the constants $D_1$ and $D_2$ in Table \ref{T3} are real numbers under the conditions $\alpha+\sqrt{6\beta}\le1+\frac{3\beta}{2}$ or $\alpha+\sqrt{6\beta}>1+\frac{3\beta}{2}$ and $\gamma-D_3>0$, at the critical points $Q_7$ and $Q_8$. Besides, the existence conditions for $Q_1$ and $Q_5$ are $\gamma\neq-\frac{1}{4}$ and $\delta\neq\frac{1}{3}$, respectively.

For the model $\Gamma_{21}$ we have four critical points $Q_1-Q_4$. The point $Q_1$ is a combination of radiation and the dark components and it corresponds to an effective fluid of radiation type. Positive energy densities at the critical points require $C_1>0$ and $\gamma<0$. The point $Q_2$ represents a combination of the dark sector and baryons and at this critical point, we always have one negative energy density and $\omega_{\textrm{eff}}=0$. For $C_4\approx0$ this point corresponds to the dominance of baryons. The points $Q_3$ and $Q_4$ are combinations of the dark sector. 
The point $Q_4$ has $\omega_{\textrm{eff}}<-1$ in the range \eqref{ranges}. For $\gamma\ll1$ we have $\omega_{\textrm{eff}}\approx \frac{\gamma}{C_4-1}$ at $Q_3$ and $\Omega_x^*\approx1$ at $Q_4$. 
\vspace{0.2cm}

For the model $\Gamma_{22}$ we have four critical points, $Q_5-Q_8$. The point $Q_5$ is a combination of radiation and the dark components and it corresponds to an effective fluid of radiation type. Positive energy densities at the critical points require $C_1>0$ and $\delta>0$ provided $3\delta<1$. The point $Q_6$ represents a combination of the dark sector and baryons and at this critical point, we always have one negative energy density and $\omega_{\textrm{eff}}=0$. For $C_4\approx0$ this point corresponds to the dominance of baryons. The points $Q_7$ and $Q_8$ are combinations of the dark sector. The point $Q_8$ has $\omega_{\textrm{eff}}<-\frac{1}{3}$ in the range \eqref{ranges} for $\delta>D_3$. For $\delta\ll1$ we have $\omega_{\textrm{eff}}\approx \frac{\delta}{1-C_4}$ at $Q_7$ and $\Omega_x^*\approx1$ at $Q_8$.

\subsection{Scenario $\Gamma_3$}
For the model $\Gamma_3$ we have, besides Eq.~\eqref{rhop}, the following set of equations,
\begin{eqnarray}
\Omega_c'&=&\frac{2 \alpha  \Omega_c-3 \beta  (\gamma+\Omega_c)-2 \Omega_c \Omega_x}{3 \beta },\\
\Omega_x'&=&\frac{2 \alpha  (\Omega_x-1)+\beta  (3 \gamma+\Omega_r-3 \Omega_x+3)-2 \Omega_x (\Omega_x-1)}{3 \beta }.
\end{eqnarray}
In the following, we describe the critical points $R_1-R_3$ in Table \ref{T4}.
In the ranges \eqref{ranges}, the constants $D_1$ and $D_2$ in Table \ref{T4} are real numbers under the conditions $\alpha+\sqrt{6\beta}\le1+\frac{3\beta}{2}$ or $\alpha+\sqrt{6\beta}>1+\frac{3\beta}{2}$ and $\gamma-D_3>0$, at the critical points $R_2$ and $R_3$.

The point $R_1$ is a combination of radiation and the dark components and it corresponds to an effective fluid of radiation type. Positive energy densities at the critical points require $C_1>0$ and $\gamma>0$.  The points $R_2$ and $R_3$ are combinations of the dark sector. 
The point $R_3$ has $\omega_{\textrm{eff}}<0$ considering the existence conditions for $D_1$ and $D_2$. For $\gamma\ll1$ we have $\omega_{\textrm{eff}}\approx \frac{\delta}{1-C_4}$ at $R_2$ and $\Omega_x^*\approx1$ at $R_3$.

\subsection{Scenario $\Gamma_4$}
For the model $\Gamma_4$ we have, besides Eq.~\eqref{rhop}, the following set of equations,
\begin{eqnarray}
\Omega_c'&=&\frac{2 \alpha  (\gamma+\Omega_c)-2 \Omega_x (\gamma+\Omega_c)-3 \beta  \Omega_c}{3 \beta },\\
 \Omega_x'&=&\frac{-2 \alpha  (\gamma-\Omega_x+1)+2 \Omega_x (\gamma-\Omega_x+1)+\beta  (\Omega_r-3 \Omega_x+3)}{3 \beta }.
\end{eqnarray}
In the following, we describe the critical points $R_4-R_6$ in Table \ref{T4}.
In the ranges \eqref{ranges} there are no additional existence conditions for critical points $R_4-R_6$.

The point $R_4$ is a combination of radiation and the dark components and it corresponds to an effective fluid of radiation type. For having positive energy densities at the critical points we must have $C_1>0$ and $\gamma<0$. The points $R_5$ and $R_6$ are combinations of the dark sector. 
The point $R_6$ has $\omega_{\textrm{eff}}<-1$ in the range \eqref{ranges}. For $\gamma\ll1$ we have $\omega_{\textrm{eff}}\approx \frac{\gamma}{C_4-1}$ at $R_5$ and $\Omega_x^*\approx1$ at $R_6$.

\subsection{Scenario $\Gamma_5$}
For the model $\Gamma_5$ we have, besides  Eq.~\eqref{rhop}, the following set of equations,
\begin{eqnarray}
\Omega_c'&=&\frac{\alpha  (2 \Omega_c-3 \gamma)+3 \gamma (\beta +\Omega_x)-\Omega_c (3 \beta +2 \Omega_x)}{3 \beta },\\
\Omega_x'&=&\frac{\alpha  (3 \gamma+2 \Omega_x-2)+\beta  (-3 \gamma+\Omega_r-3 \Omega_x+3)+\Omega_x (-3 \gamma-2 \Omega_x+2)}{3 \beta }.
\end{eqnarray}
In the following, we describe the critical points $R_7-R_9$ in Table \ref{T4}. 
Notice that the constants $E_2$ and $E_3$ are real numbers in the range \eqref{ranges}.

The point $R_7$ is a combination of radiation and the dark components and it corresponds to an effective fluid of radiation type. Positive energy densities at the critical points require $C_1>0$ and $\gamma>0$.  The points $R_8$ and $R_9$ are combinations of the dark sector. The point $R_9$ has $\omega_{\textrm{eff}}<-\frac{1}{3}$ in the range \eqref{ranges}. For $\gamma\ll1$ we have $\omega_{\textrm{eff}}\approx \frac{\gamma}{2(1-C_4)}$ at $R_8$ and $\Omega_x^*\approx1$ at $R_9$.

\begin{table*}[ht!]
\begin{center}\begin{minipage}{\textwidth}
\centering
\begin{tabular}{ccccc}\hline

&Critical Points & $\omega_{\rm eff}$ & Stability Conditions & Classification\\\hline
\multicolumn{5}{c}{$ \Gamma_{3}$}\\\hline
\multirow{3}{*}{$R_1$}&\multirow{3}{*}{$\displaystyle\{1-C_1-3\gamma,3\gamma,C_1\}$} & 
\multirow{3}{*}{$\displaystyle\frac{1}{3}$}& $1-C_4\ge\sqrt{6\beta}$, $\gamma<\frac{1}{3}(1-C_1)$ &unstable\\
&&&$1-C_4<\sqrt{6\beta}$, $D_3\le\gamma<\frac{1}{3}(1-C_1)$ &unstable\\
&&&$\gamma>\frac{1}{3}(1-C_1)$&saddle\\[3pt] \cline{1-5}
\multirow{4}{*}{$R_2$}&\multirow{4}{*}{$\displaystyle\left\{0,\frac{1}{2}(1-D_1),\frac{1}{2}(1+D_1)\right\}$ } &
\multirow{4}{*}{$\displaystyle\frac{D_2-1}{3\beta}$}& $\gamma>\frac{1}{3}(C_1-1)$& unstable\\
&&&$1-C_4\ge\sqrt{6\beta}$, $\gamma<0$&saddle\\
&&&$1-C_4<\sqrt{6\beta}$, $D_3<\gamma<0$&saddle\\
&&&$\gamma<\frac{1}{3}(C_1-1)$&saddle\\[3pt] \cline{1-5}
\multirow{2}{*}{$R_3$}&\multirow{2}{*}{$\displaystyle\left\{0,\frac{1}{2}(1-D_2),\frac{1}{2}(1+D_2)\right\}$} &
\multirow{2}{*}{$\displaystyle\frac{D_1-1}{3\beta}$}& $1-C_4\ge \sqrt{6\beta}$& stable\\
&&&$1-C_4< \sqrt{6\beta}$, $\gamma>D_3$& stable\\\hline\cline{1-5}
\multicolumn{5}{c}{$ \Gamma_{4}$}\\\hline
\multirow{5}{*}{$R_4$}&\multirow{5}{*}{$\left\{1-C_1+4\gamma,-4\gamma,C_1\right\}$} & 
\multirow{5}{*}{$\displaystyle\frac{1}{3}$} & $0<\beta<\frac{2}{3}$, $E_1<\gamma\le-\frac{\beta}{2}$, $\alpha<\tilde{C_5}$& unstable\\
&& &$0<\beta<\frac{2}{3}$, $\gamma>-\frac{\beta}{2}$&unstable\\ 
&& &$\beta\ge\frac{2}{3}$, $E_1<\gamma\le-\frac{\beta}{2}$, $\alpha<\tilde{C_5}$&unstable\\ 
&& &$\beta\ge\frac{2}{3}$, $\gamma\ge -\frac{\beta}{2}$&unstable\\ 
&& &$\gamma\le E_1$ or $E_1<\gamma<-\frac{\beta}{2}$, $\gamma>C_5$&saddle\\ [3pt]\hline
\multirow{6}{*}{$R_5$}&\multirow{6}{*}{$\displaystyle\left\{0,\frac{1}{2}(1+C_6),\frac{1}{2}(1-C_6)\right\}$} & 
\multirow{6}{*}{$\displaystyle-\frac{1+C_7+2\gamma}{3\beta}$} & $\gamma\le E_1$ or $E_1<\gamma<-\frac{\beta}{2}$, $\alpha>C_5$& unstable\\
&& &$\gamma>0$&saddle\\
&& &$\beta<\frac{2}{3}$, $E_1<\gamma<-\frac{\beta}{2}$, $\alpha<C_5$&saddle\\
&& &$\beta<\frac{2}{3}$, $-\frac{\beta}{2}<\gamma<0$&saddle\\
&& &$\beta\ge\frac{2}{3}$, $E_1<\gamma<-\frac{\beta}{2}$, $\alpha<C_5$&saddle\\
&& &$\beta\ge\frac{2}{3}$, $-\frac{\beta}{2}\le\gamma<0$&saddle\\[3pt] \cline{1-5}
$R_6$&$\displaystyle\left\{0,\frac{1}{2}(1+C_7),\frac{1}{2}(1-C_7)\right\}$ & $\displaystyle-\frac{1+C_6+2\gamma}{3\beta}$& -&stable\\\hline\cline{1-5}

\multicolumn{5}{c}{$ \Gamma_{5}$}\\\hline
\multirow{4}{*}{$R_7$}&\multirow{4}{*}{$\displaystyle\left\{1-C_1-3\gamma,3\gamma,C_1\right\}$} & 
\multirow{4}{*}{$\displaystyle\frac{1}{3}$} & $\gamma\le\frac{2\beta}{3}$ &unstable\\
&&&$\frac{2\beta}{3}<\gamma<-\frac{4E_1}{3}$, $\alpha<1+2\beta-3\gamma$&unstable\\
&&&$\frac{2\beta}{3}<\gamma\le-\frac{4E_1}{3}$, $\alpha>1+2\beta-3\gamma$&saddle \\
&&&$\gamma>-\frac{4E_1}{3}$&saddle \\[3pt]\cline{1-5}

\multirow{4}{*}{$R_8$}&\multirow{4}{*}{$\displaystyle\left\{0,\frac{1}{2}(1-E_2+\frac{3}{2}\gamma),\frac{1}{2}(1+E_2-\frac{3}{2}\gamma)\right\}$} & 
\multirow{4}{*}{$\displaystyle-\frac{1-E_3-\frac{3}{2}\gamma}{3\beta}$} & $\frac{2\beta}{3}<\gamma\le-\frac{4E_1}{3}$, $\alpha>1-3\gamma+2\beta$&unstable\\
&&&$\gamma>-\frac{4E_1}{3}$&unstable\\
&&&$\gamma<0$ or $0<\gamma\le \frac{2\beta}{3}$&saddle\\
&&&$\frac{2\beta}{3}<\gamma<-\frac{4E_1}{3}$, $\alpha<1-3\gamma+2\beta$&saddle\\ [3pt]\hline

$R_9$&$\displaystyle\left\{0,\frac{1}{2}(1-E_3+\frac{3}{2}\gamma),\frac{1}{2}(1+E_3-\frac{3}{2}\gamma)\right\}$ &
$\displaystyle-\frac{1-E_2-\frac{3}{2}\gamma}{3\beta}$& -& stable\\
\cline{1-5}

\end{tabular}
\end{minipage}
\end{center}
\caption{\label{T4} Description of the critical points $(\Omega_{r}^*,\Omega_{c}^*,\Omega_{x}^*)$ for scenario $\Gamma_{2}$. The ranges in \eqref{ranges} are considered as global constraints. To improve visualization we use the constants $E_1-E_3$ defined in \eqref{Const3}.}

Summarizing the dynamical system results of MHR-IDE models, $\Gamma_1-\Gamma_5$, we can indicate that the radiation epoch is always modified for these models, but by considering $\alpha-2\beta\ll1$ and a small interaction, the modifications are ameliorated. A critical point corresponding to an attractor is present in each model and it represents an accelerated phase. There is also a saddle point corresponding to a combination of the dark sector. 

\end{table*}

\section{Observational analysis and model selection}\label{sec4}

In order to constrain the MHR-IDE models, we use different data sources such as type Ia supernovae from the Pantheon sample \cite{bib3}, cosmic chronometers \cite{bib1}, baryon acoustic oscillations \cite{bib5} \cite{bib6}
\cite{bib7} and the position of the angular scale of the sound horizon at last scattering \cite{bib11}. We briefly present each data set below.

\subsection{Supernovae type Ia}

We use the Pantheon sample, which contains a set of 1048 spectroscopically confirmed SNe Ia in the redshift range $0.01<z<2.3$ \cite{bib3}. This data set contains measurements of the apparent magnitude $m_b$, which is related to the distance modulus by $\mu=m_b - M_b$, where the absolute magnitude $M_b$ is a nuisance parameter. In terms of the luminosity distance, $d_L$, the distance modulus is defined as,
\begin{equation}
    \mu = 5 \log{d_L} + 25,
\end{equation}
where
\begin{equation}
d_L = (1+z)\int_{0}^{z} \frac{dz'}{H(z')}.
\end{equation}
is measured in Mpc.

\subsection{Cosmic chronometers}

This data set corresponds to 24 measurements of the expansion rate at different redshift values. These cosmic chronometers are obtained from the differential age method \cite{bib4}, from which we only consider data at redshift $z<1.2$ (see Refs.~\cite{bib1} and \cite{bib2}). Notice that these data constitute the only method that provides cosmology-independent, direct measurements of the expansion history of the universe \cite{bib2}.

\subsection{Baryon acoustic oscillations (BAO)}

We use three isotropic BAO measurements from 6dFGS \cite{bib5}, MGS \cite{bib6} and eBOSS \cite{bib7} and three data points from the BOSS DR12 \cite{bib8}, corresponding to BAO anisotropic measurements. All of these BAO measurements are given in terms of $D_V(z)/r_s$, $D_M(z)/r_s$ and $D_H(z)/r_s$, where $D_V$ is a combination of the line-of-sight and transverse distance scales defined in Ref. \cite{bib9}, $D_M(z)$ is the comoving angular diameter distance, which is related to the physical angular diameter distance by $D_M(z) = (1+z)D_A(z)$ and $D_H = c/H(z)$ is the Hubble distance. We define $D_V(z)$ and $D_A(z)$ as
\begin{equation}
    D_V(z) =\left( D_M^2(z)\frac{z}{H(z)} \right)^{1/3}
\end{equation}
\begin{equation}
    D_A(z) = \frac{1}{(1+z)}\int_{0}^z \frac{dz'}{H(z')} \ ,
\end{equation}
The standard ruler length $r_s$ is the comoving size of the sound horizon at the drag epoch, defined as
\begin{equation}
    r_s = \int_{z_d}^\infty \frac{c_s\, dz}{H(z)},
\end{equation}
where  $c_s = \frac{1}{\sqrt{3(1+\mathcal{R})}}$ is the sound speed in the photon-baryon fluid,  $\displaystyle\mathcal{R} = \frac{3\Omega_{b0}}{4\Omega_{\gamma 0}(1+z)}$ \cite{bib10} and $z_d$ the redshift at the drag epoch. 
  
\subsection{Cosmic microwave background}
We consider one data point corresponding to the position of the angular scale of the sound horizon at last scattering, as background data coming from the early universe physics,
\begin{equation}
\textit{l}_a = \frac{\pi D_M(z_*)}{r_s(z_*)},
\end{equation}
where $z_* = 1089.80$, according with Planck's 2018 \cite{bib11}. We compare our calculated value with the one reported by the Planck collaboration in 2015, $l_a = 301.63 \pm 0.15$ \cite{Planck:2015bue}. 

\subsection{Bayesian model selection}

The evaluation of a model's performance in light of the data is based on the \textit{Bayesian evidence} \cite{bib12}. This is the normalization integral on the right-hand-side of Bayes' theorem, which is related to the posterior probability $P$ for a set of parameters  $\Theta$, given the data $\mathcal{D}$, described by a model $\mathcal{M}$,
\begin{equation}
    P(\Theta \vert \mathcal{D},\mathcal{M}) = \frac{\mathcal{L}(\mathcal{D}\vert \Theta,\mathcal{M})\mathcal{P}(\Theta \vert \mathcal{M})}{\mathcal{E}(\mathcal{D}\vert \mathcal{M})},
\end{equation}
where $\mathcal{L},\mathcal{P}$ and $\mathcal{E}$ are the likelihood, prior distribution, and evidence, respectively. We can write the evidence for a continuous parameter space $\Omega_{\mathcal{M}}$ as
\begin{equation}
\mathcal{E}(\mathcal{D} \vert \mathcal{M}) = \int_{\Omega_{\mathcal{M}}} \mathcal{L}(\mathcal{D}\vert \Theta,\mathcal{M})\mathcal{P}(\Theta \vert \mathcal{M}).
\end{equation}

In order to compare the performance of different models given a dataset, we use the Bayes' factor defined as the ratio of the evidence of models $\mathcal{M}_0$ and $\mathcal{M}_1$ as:
\begin{equation}
    B_{01} = \frac{\mathcal{E}(\mathcal{D} \vert \mathcal{M}_0)}{\mathcal{E}(\mathcal{D} \vert \mathcal{M}_1)},
\end{equation}
which we will use to interpret the strength of the evidence through Jeffrey's scale given in Table \ref{tab2}  and referenced in \cite{bib12}. This is an empirically calibrated scale, representing weak, moderate, or strong evidence. In our work, we consider the $\Lambda \text{CDM}$  model as reference model ($\mathcal{M}_{1}$), therefore if $\ln B_{01}<0$ we will have evidence in favor of $\Lambda \text{CDM}$, on the other hand, if $\ln B_{01}>0$ the evidence will favor the MHR-IDE model.

\begin{table}[htb]
\centering
\begin{tabular}{ c  c  c  c  c  }
\hline
$\vert \ln{B_{01}} \vert$ &  Odds & Probability & Strength of evidence  \\ \hline
 $<1.0$ & $ \lesssim 3:1$  & $<0.750$  & Inconclusive \\
 $1.0$& $ \sim 3:1$  & $0.750$  & Weak    \\
 $2.5$& $ \sim 12:1$  & $0.923$  & Moderate \\
 $5.0$& $ \sim 150:1$ & $0.993$  & Strong  \\
 \hline
 &&&
\end{tabular}
\caption{The Jeffreys' scale, empirical measure for interpreting the evidence in comparing two models $\mathcal{M}_0$ and $\mathcal{M}_1$ as presented in reference \cite{bib12}. The probability column refers to the posterior probability of the favored model, assuming both models are equally likely and fill the entire model space. }\label{tab2}

\end{table}

In general, for a set of measurements contained in a vector $\mathcal{S}$, we have the $\chi^2$ function defined as:
\begin{equation}
    \chi_{\mathcal{S}}^2 = [\mathcal{S}^{\textrm obs}-\mathcal{S}^{\textrm th}]^T \mathcal{C}^{-1}[\mathcal{S}^{\textrm obs}-\mathcal{S}^{\textrm th}]
\end{equation}
where $\mathcal{S}^{obs}$ represents the measured value, $\mathcal{S}^{th}$ is the theoretical value computed
assuming a model with parameters $\Theta$ and $\mathcal{C}$ corresponds to the covariance matrix of the measurements contained in the vector $\mathcal{S}^{obs}$. In our case the values in $\mathcal{S}^{th}$ represent the functions $\mu(z)$, $H(z)$ and $l_a(z_*)$ for SNe-Ia, CC and CMB data, respectively, and $D_V(z)/r_s$, $D_M(z)/r_s$ or $D_H(z)/r_s$ for BAO data.

The analyses for all samples were performed assuming a multivariate Gaussian likelihood of the form
\begin{equation}
    \mathcal{L}(\mathcal{D}|\Theta) = \exp\left[-\frac{\chi^2(\mathcal{D}|\Theta)}{2}\right].
\end{equation}
To find the best-fit model parameters  we perform a joint analysis including all the data, we use the overall $\chi^2$ function defined as

\begin{equation}
    \chi^2 = \chi^2_{\textrm{SNe-Ia}} + \chi^2_{\textrm{CC}} + \chi^2_{\textrm{BAO}} + \chi^2_{\textrm{CMB}}.
\end{equation}

To calculate the evidence and estimate the cosmological parameters we use the MULTINEST algorithm \cite{bib13,bib14}, setting a tolerance of $0.01$ as convergence criterion and working with a set of $1000$ live points to improve the precision in the estimation of the evidence.


\section{ Analysis and results}\label{sec5}

We perform a Bayesian comparison of models $\Gamma_{11}$, $\Gamma_{12}$, $\Gamma_{13}$ and $\Gamma_{14}$ representing the variations of $\Gamma_1$ for the cases where $\delta=0$, $\gamma=0$, $\delta =\gamma$ and $\delta \neq \gamma$, respectively. Similarly, $\Gamma_{21}$, $\Gamma_{22}$, $\Gamma_{23}$ and $\Gamma_{24}$ are the variations of $\Gamma_2$ for the same cases. In this work, we used the priors shown in Table \ref{tab3} and the combination of all the background data displayed in section \ref{sec4}, where we have considered two different approaches in using the data whose purpose is to clarify whether there is a noticeable impact on our results when considering different priors for the Hubble parameter $h$. Considering a Gaussian prior is equivalent to including the local measure of $H_0$ \cite{Riess:2021jrx} in the data set. Notice that for the parameters $\gamma$ and $\delta$, we have explored the analytical solutions of our models inside the parameter space, where we have found an interval between -0.1 and 0.1 for these parameters, in which our functions are well defined. 

We consider a joint analysis with the dataset Pantheon + CC + BAO + CMB, as described in section \ref{sec4}, which is studied in two different scenarios distinguished by the assigned prior for the $h$ parameter (see Tables \ref{tab4} and \ref{tab5}). Our main results are summarized in Tables \ref{tab4}$-$\ref{tab6}.

\begin{table}[htb!]
\centering
\begin{tabular}{c c c}\hline
Parameter & Prior & Ref. \\ \hline 
$h$ & Gaussian : (0.7403, 0.0142)&  \cite{bib19}   \\
& Uniform: (0.6, 0.8) & \ \ \ \ - \\ 
$\Omega_c$ & Uniform: (0, 1) &  \ \ \ \ - \\
$\alpha$ & Uniform: (0, 1) & \cite{bib16} \cite{bib17}  \\
$\beta$ & Uniform: (0, 1) & \cite{bib16} \cite{bib17}   \\
$\delta$ & Uniform: (-0.1, 0.1) &   \\
$\gamma$ & Uniform: (-0.1, 0.1) &  \\
$M_B$ & Uniform: (-20, -18) & \cite{bib18}  \\
\hline
\end{tabular}%
\caption{Priors on the free parameters of the MHR-IDE models. For the Gaussian prior we inform ($\mu$, $\sigma^2$) and for the uniform prior, $(a,b)$ represents $a < x < b$.}\label{tab3}
\end{table}

In our analysis, we fix the following parameters, under the assumption that the variation in the radiation and baryonic components is not significant,
\begin{eqnarray}
\Omega_{r0}=\left(1+\frac{7}{8}\left(\frac{4}{11}\right)^{\frac{4}  {3}}N_{\textrm{eff}}\right)\Omega_{\gamma0},\quad \Omega_{b0}=0.02235,
\end{eqnarray}
where $N_{\textrm{eff}} = 3.046$ [68.], $\Omega_{\gamma0} = 2.469 \times 10^{-5}$ and $\Omega_{b0}$ [69.] correspond to the effective number of neutrinos, the photon density parameter and the baryon density parameter, respectively.

Tables \ref{tab4} and \ref{tab5} present the best-fit parameters with their associated $1\sigma$ error for the MHR-IDE models studied in this work, for a Gaussian and a Uniform prior for the $h$ parameter, respectively. Table \ref{tab6} shows the logarithm of the Bayesian evidence and the logarithm of the Bayes factor which is interpreted in terms of the Jeffreys' scale \eqref{tab2}. As a comparison, in Tables \ref{tab4} $-$ \ref{tab6} we also show the results for HRDE and MHRDE models.

We note that the strength of the Bayesian evidence for the Gaussian prior is weak/moderate for the MHR-IDE models in Table \ref{tab1}, while considering the uniform prior the evidence is strong for all models in Table \ref{tab1}, i.e., when we consider a uniform prior for the $h$ parameter, the evidence supports the $\Lambda$CDM model. However, independent of the prior used, the evidence favors $\Lambda$CDM for all the MHR-IDE studied.

\begin{figure}[htb]
\centering
\includegraphics[width=0.45\textwidth]{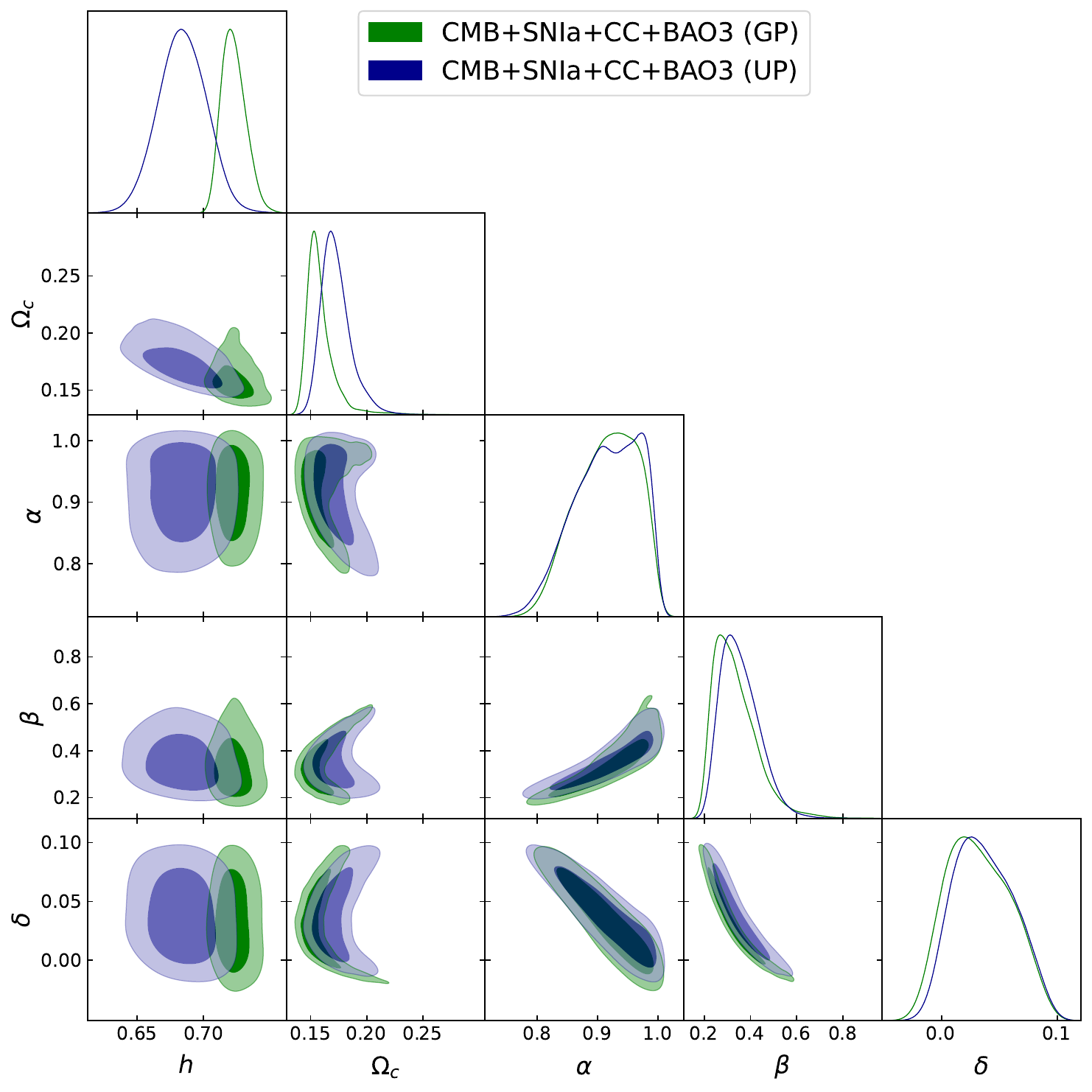}  \includegraphics[width=0.45\textwidth]{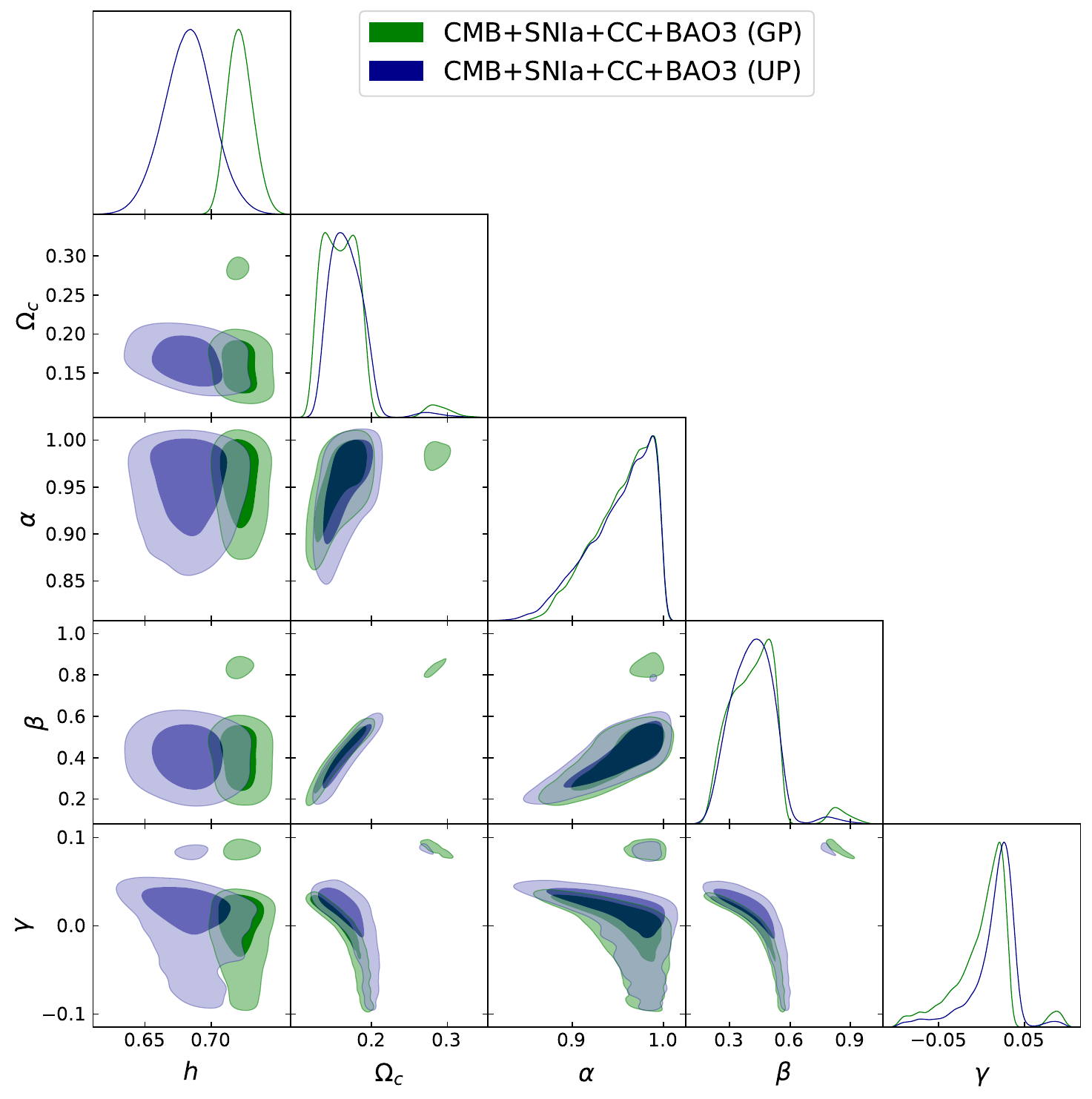}
\caption{Contour plots with $1\sigma$ and $2\sigma$ regions are shown for $\Gamma_{11}$ in the left panel and $\Gamma_{12}$ in the right panel. We considered the full joint
analysis with Pantheon + CC + BAO + CMB. \label{F1}}
\end{figure}

\begin{figure}[htb]
\centering
\includegraphics[width=0.45\textwidth]{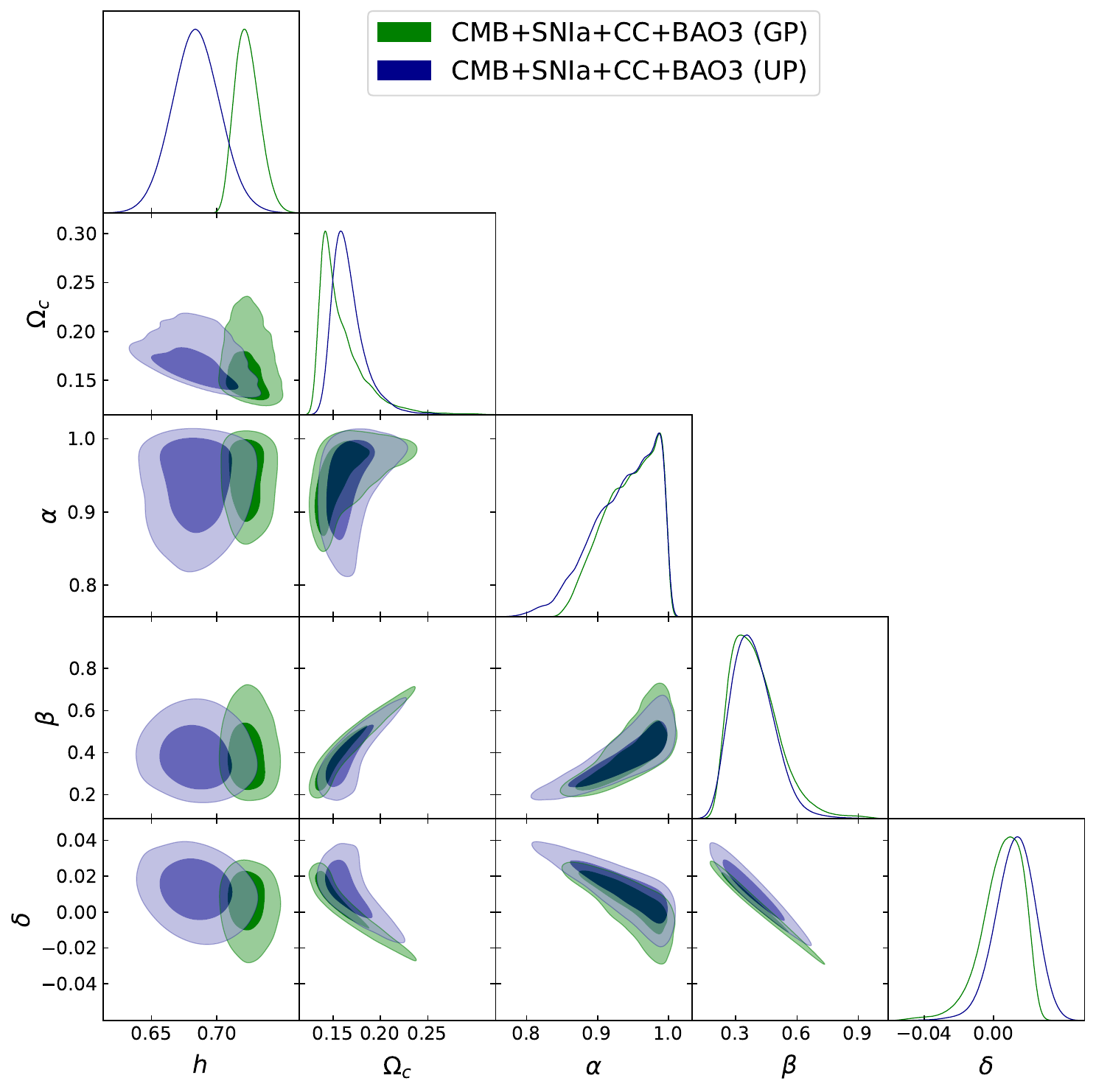}
\includegraphics[width=0.45\textwidth]{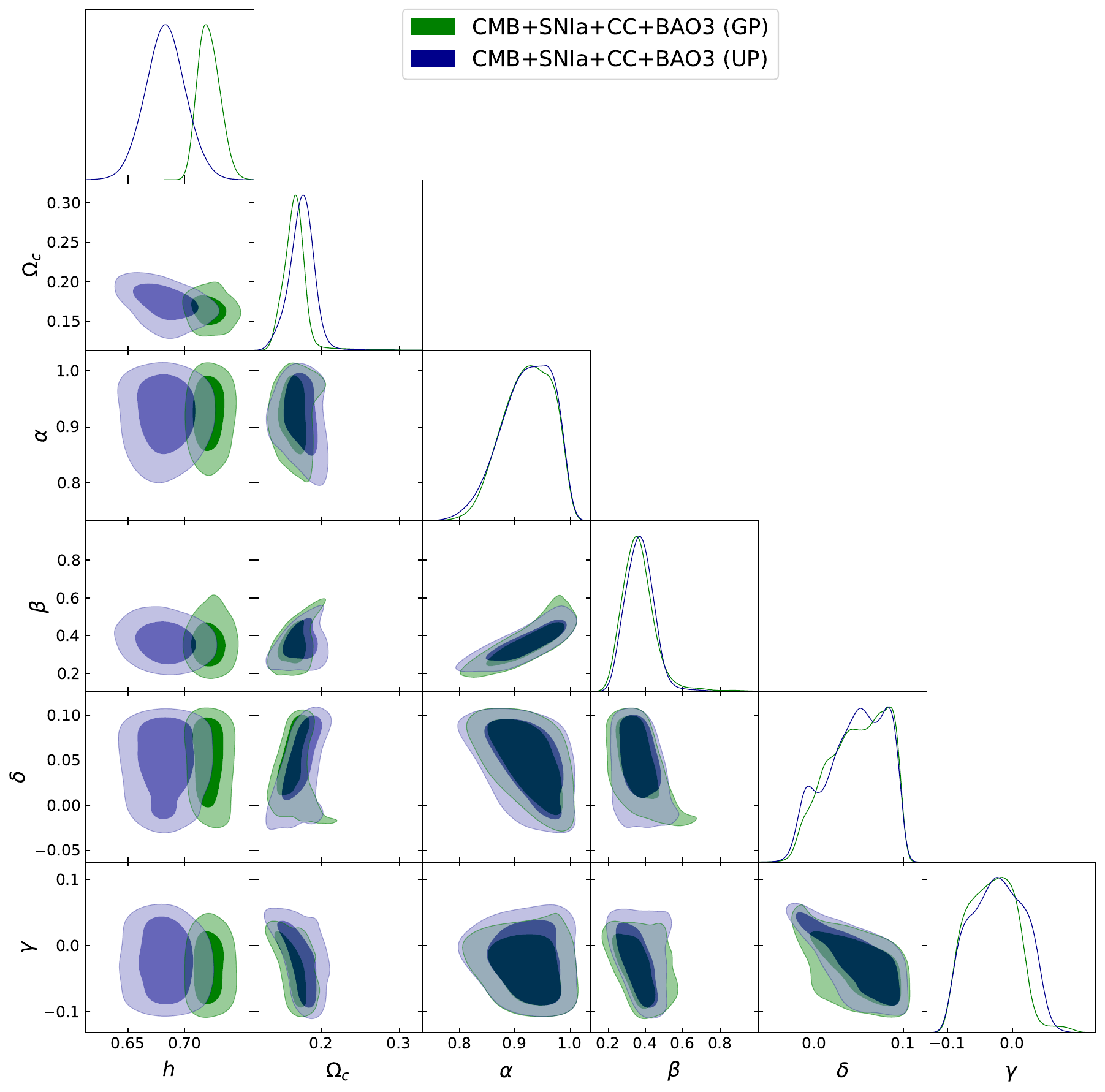}
\caption{Contour plots with $1\sigma$ and $2\sigma$ regions are shown for $\Gamma_{13}$ in the left panel and $\Gamma_{14}$ in the right panel. We considered the full joint
analysis with Pantheon + CC + BAO + CMB. \label{F2}}
\end{figure}

\begin{figure}[htb]
\centering
\includegraphics[width=0.45\textwidth]{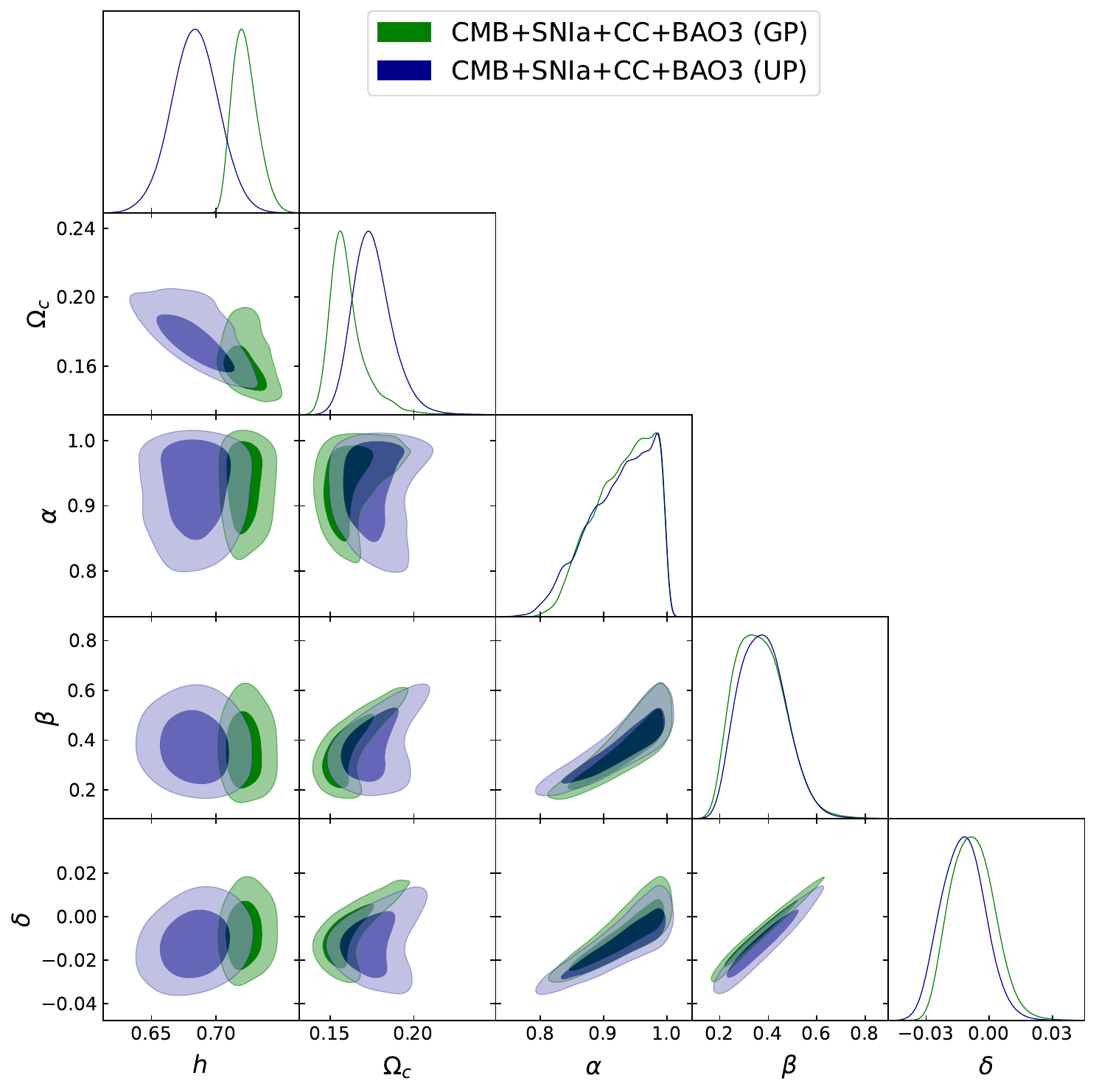}
\includegraphics[width=0.45\textwidth]{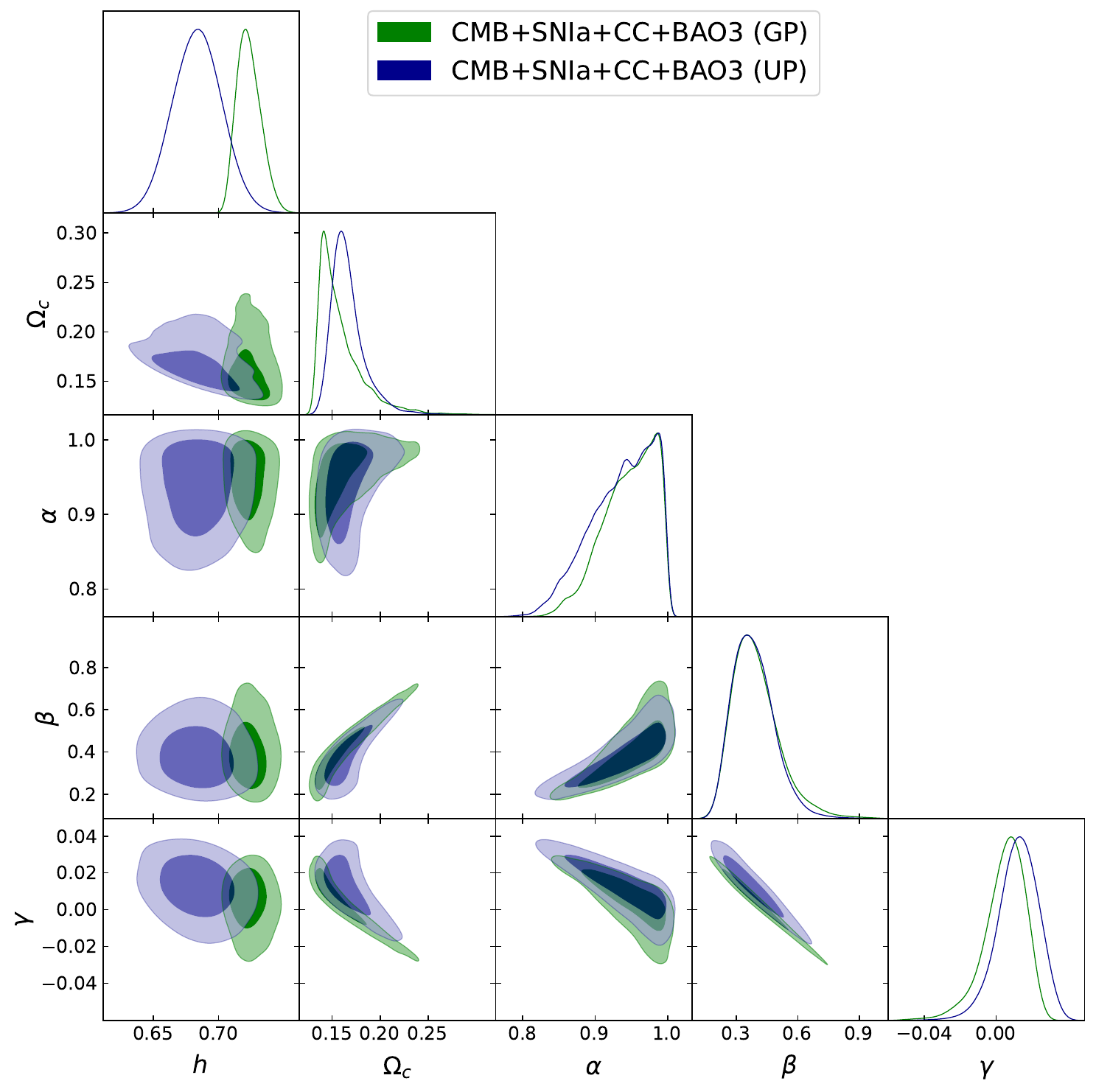}
\caption{Contour plots with $1\sigma$ and $2\sigma$ regions are shown for $\Gamma_{21}$ in the left panel and $\Gamma_{22}$ in the right panel. We considered the full joint
analysis with Pantheon + CC + BAO + CMB. \label{F3}}
\end{figure}

\begin{figure}[htb]
 \centering
\includegraphics[width=0.45\textwidth]{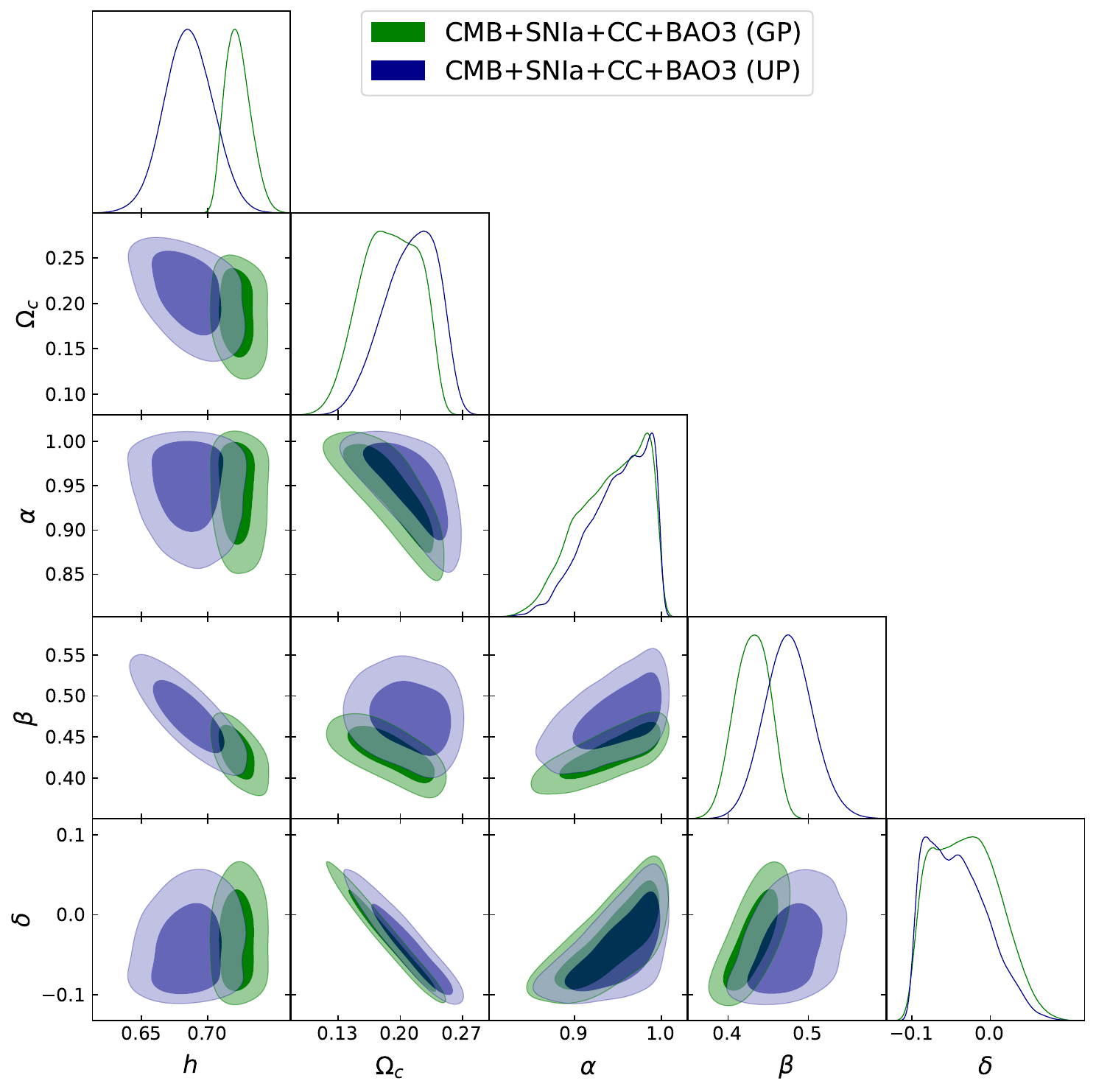}
\includegraphics[width=0.45\textwidth]{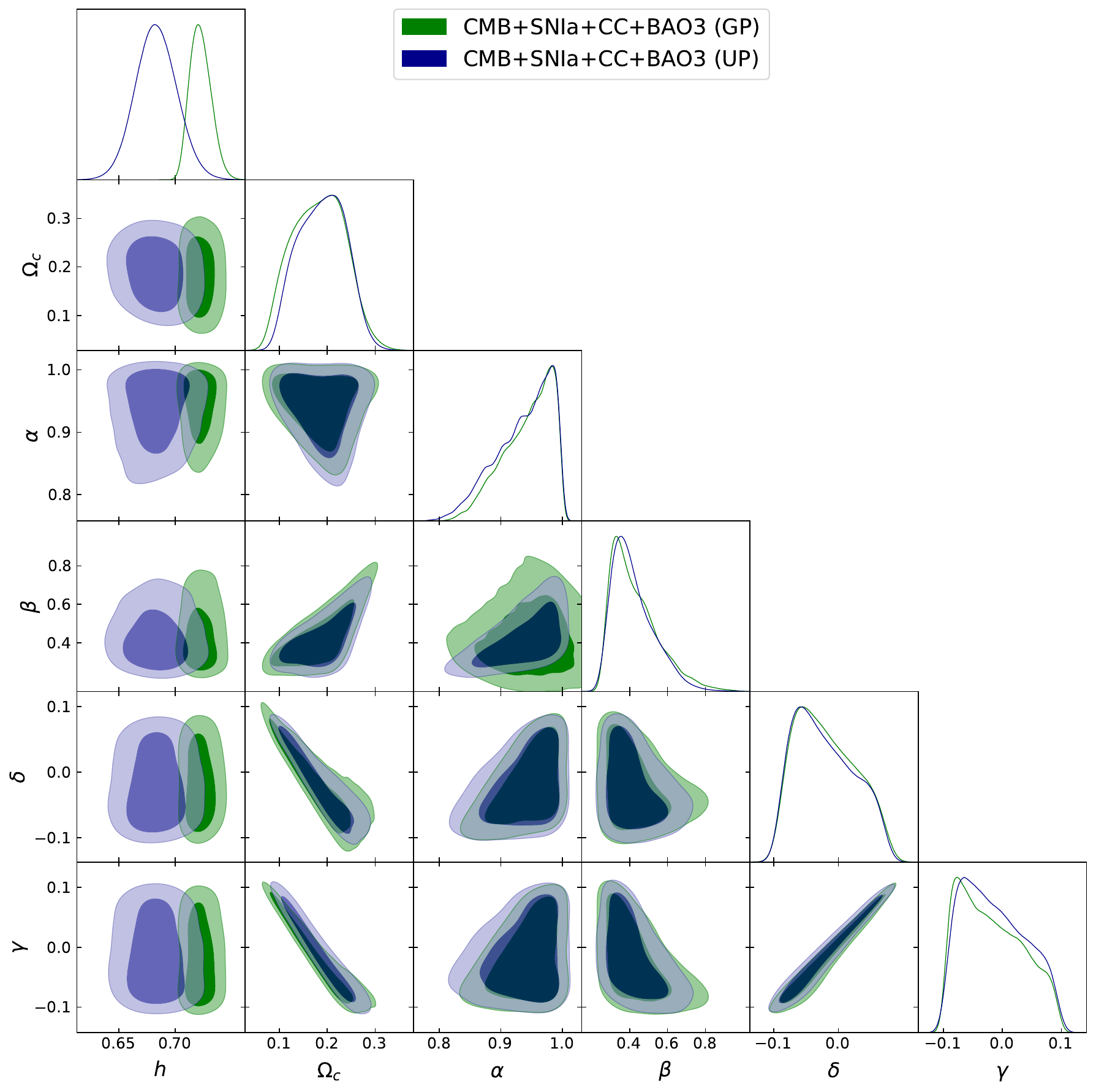}
\caption{Contour plots with $1\sigma$ and $2\sigma$ regions are shown for $\Gamma_{23}$ in the left panel and $\Gamma_{24}$ in the right panel. We considered the full joint
analysis with Pantheon + CC + BAO + CMB. \label{F4}}
\end{figure}

\begin{figure}[htb]
\centering
\includegraphics[width=0.45\textwidth]{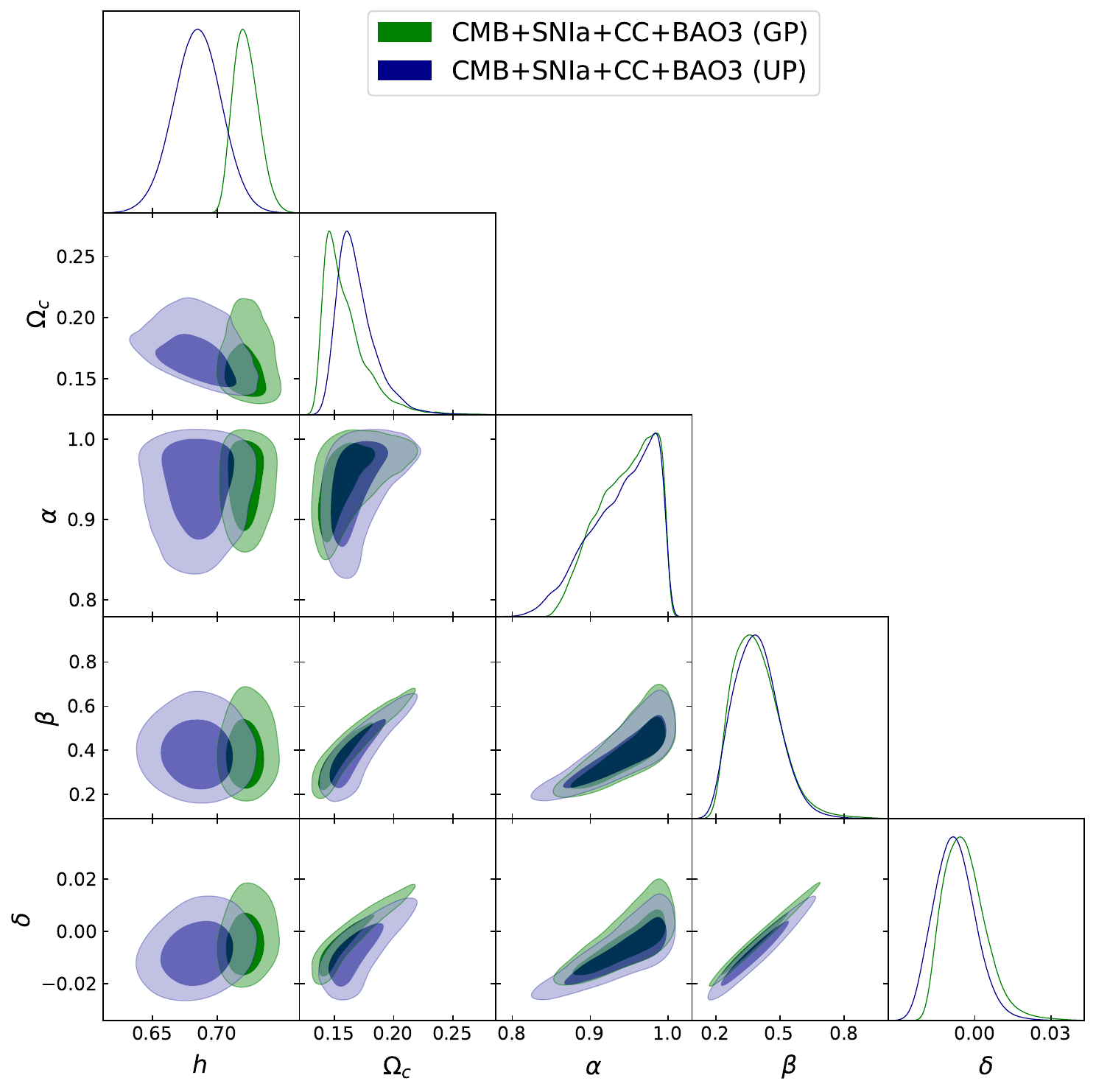}
\includegraphics[width=0.45\textwidth]{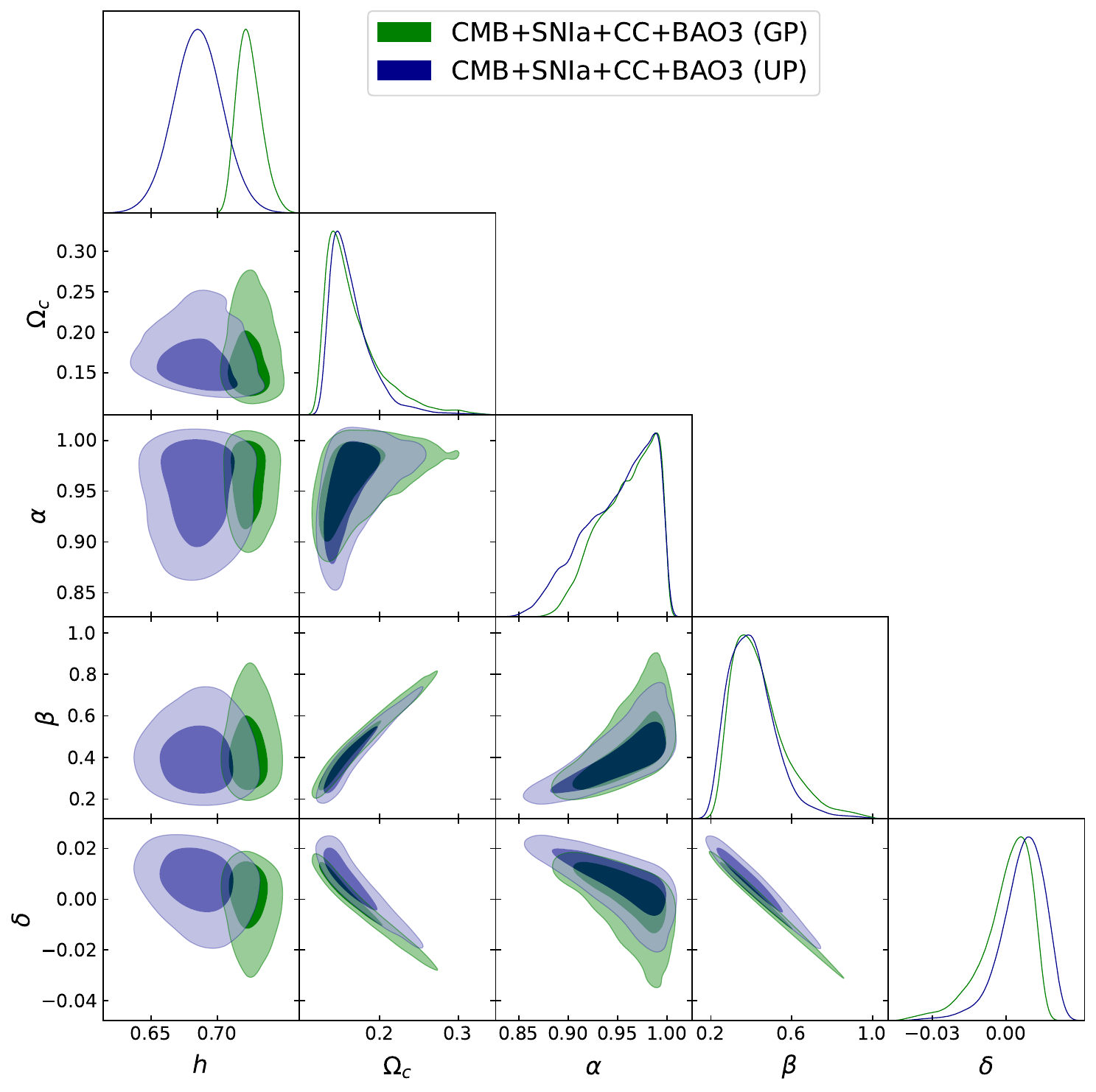}
\caption{Contour plots with $1\sigma$ and $2\sigma$ regions are shown for $\Gamma_{3}$ in the left panel and $\Gamma_{4}$ in the right panel. We considered the full joint
analysis with Pantheon + CC + BAO + CMB. \label{F5}}
\end{figure}

\begin{figure}[htb]
\centering
\includegraphics[width=0.45\textwidth]{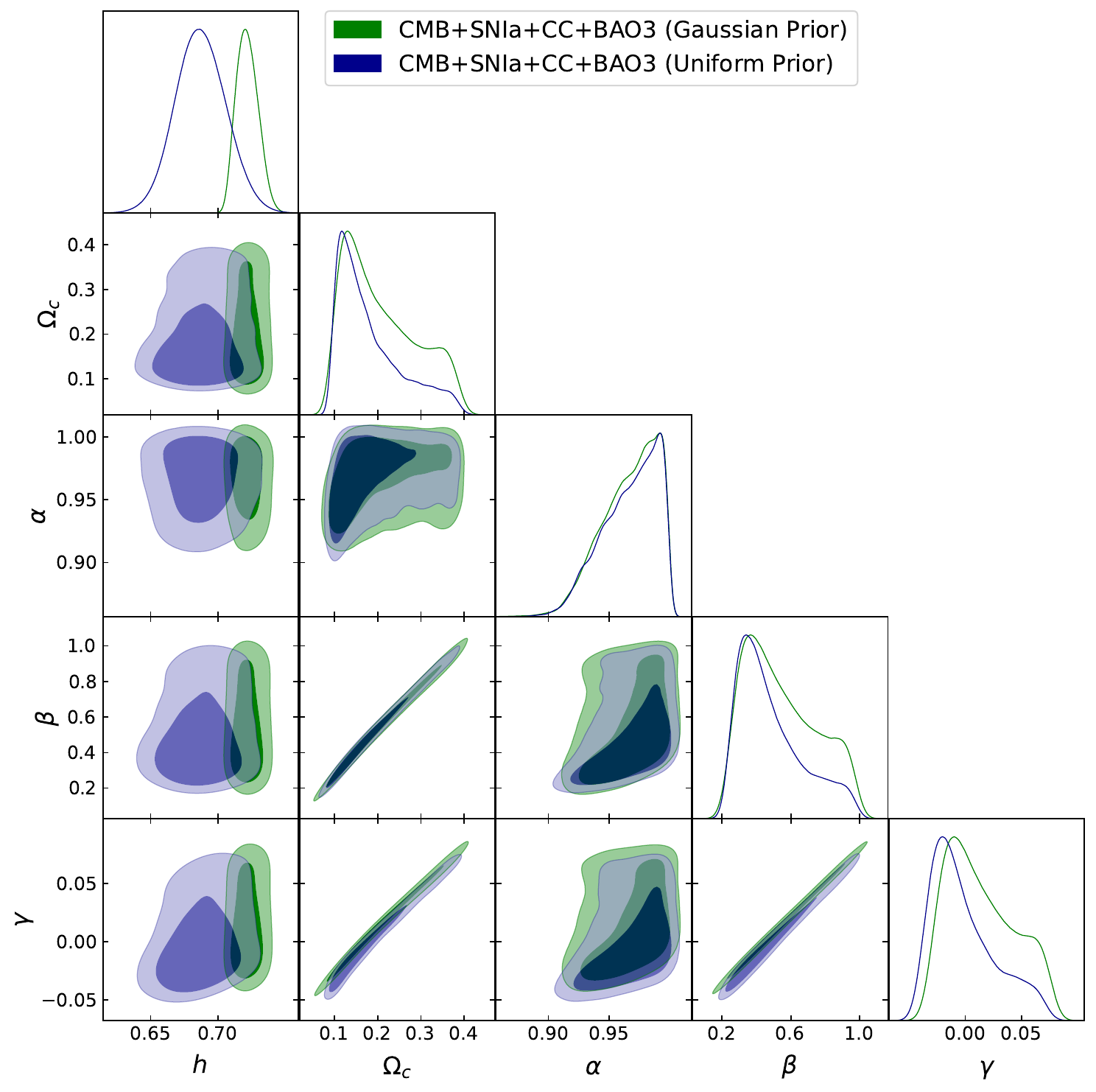}
\caption{Contour plot with $1\sigma$ and $2\sigma$ regions is shown for $\Gamma_{5}$. We considered the full joint analysis with Pantheon + CC + BAO + CMB.\label{F6}}
\end{figure}

Figures \ref{F1}$-$\ref{F6} show the contour plots for models $\Gamma_1-\Gamma_5$ with $1\sigma$  and $2\sigma$ confidence levels, where we have considered the joint analysis for each distribution of parameter $h$, in green for the case with Gaussian prior and blue for the uniform prior. In general, in all the models but $\Gamma_{23}$, the parameters associated with interaction and holography coincide inside the $1\sigma$ region when the prior for $h$ is changed. On the other hand, the parameters $\alpha$ and $\beta$ in general (all scenarios but $\Gamma_{23}$, $\Gamma_{24}$ and $\Gamma_5$) exhibit some degree of correlation approaching $\alpha\approx2\beta$. Finally, all the models in Tables \ref{tab4} and \ref{tab5}, but $\Gamma_{11}$, are compatible with null interaction.

In references \cite{Li:2009bn,delCampo:2011jp,Wang:2010kwa,Fu:2011ab,Feng:2016djj,George:2019vko,Akhlaghi:2018knk,DAgostino:2019wko} the performance of holographic dark energy models in fitting cosmological data has been assessed, compared to the $\Lambda$CDM model, where several criteria have been used, $\chi^2/dof$, AIC and BIC \cite{Arevalo:2016epc} and Bayesian evidence \cite{Cid:2020kpp}. To our knowledge, no evidence in favor of any interacting holographic dark energy scenarios has been found.

We notice that for a Gaussian prior on $h$, the MHRDE model presents inconclusive evidence. In this sense, the authors of Ref.~\cite{Cid:2020kpp} previously found weak evidence in favor of $\Lambda$CDM when they compared the MHRDE for the same dataset, however, they use different mean for the Gaussian prior on $h$.

Finally, the authors of Ref.~\cite{Li:2009bn} studied the HRDE model, finding evidence against the HRDE model compared with $\Lambda$CDM. In addition, in Ref.~\cite{Akhlaghi:2018knk}, the HRDE and MHRDE models are studied using expansion and growth data of structures, through the BIC information criteria. The authors find strong indications against the HRDE models when compared to $\Lambda$CDM.

\begin{table}
\centering
\resizebox{15cm}{!}{%
\begin{tabular}{ c  c  c  c  c  c  c  }
\hline
Model & $h$ & $\Omega_{c}$ & $\alpha$ & $\beta$ & $\gamma$ & $\delta$ \\ \hline
$\Lambda\text{CDM}$ & 0.6942 $\pm$ 0.0047 & 0.2411 $\pm$ 0.0058& - & -  & - & -  \\
HRDE & 0.7043 $\pm$ 0.0054 & 0.1745 $\pm$ 0.0052& - & 0.471 $\pm$ 0.012  & - & -  \\
MHRDE & 0.7211 $\pm$ 0.0081 & 0.1651 $\pm$ 0.0052 & 0.970 $\pm$ 0.020 & 0.445 $\pm$ 0.013  & - & -  \\
$\Gamma_{11}$ & 0.7205 $\pm$ 0.0080 & 0.1603 $\pm$ 0.0158 & 0.915 $\pm$ 0.051& 0.338 $\pm$ 0.101  & 0.033 $\pm$ 0.028 & -  \\
$\Gamma_{12}$& 0.7204 $\pm$ 0.0080 &0.1612 $\pm$ 0.0328 & 0.952 $\pm$ 0.033 & 0.412 $\pm$ 0.130  & - & 0.004 $\pm$ 0.033  \\
$\Gamma_{13}$ & 0.7204 $\pm$ 0.0083 & 0.1573 $\pm$ 0.0244 & 0.940 $\pm$ 0.039 & 0.378 $\pm$ 0.120 & 0.007 $\pm$ 0.012  & 0.007 $\pm$ 0.012 \\
$\Gamma_{14}$& 0.7203 $\pm$ 0.0075 & 0.1663 $\pm$ 0.0173 & 0.923 $\pm$ 0.046 & 0.361 $\pm$ 0.094 & -0.034 $\pm$ 0.036& 0.051 $\pm$ 0.031  \\
$\Gamma_{21}$ & 0.7205 $\pm$ 0.0082  & 0.1609 $\pm$ 0.0114 & 0.931 $\pm$ 0.044  & 0.364 $\pm$ 0.100   & -0.008$\pm$ 0.010 & - \\
$\Gamma_{22}$ & 0.7213   $\pm$ 0.0077 & 0.1588  $\pm$ 0.0251 & 0.944 $\pm$ 0.037 & 0.390  $\pm$ 0.118  & - & 0.006  $\pm$ 0.012  \\
$\Gamma_{23}$ &  0.7213 $\pm$ 0.0079  & 0.1898 $\pm$ 0.0309 &  0.943 $\pm$ 0.037 &  0.432 $\pm$ 0.020 & -0.032 $\pm$ 0.041  & -0.032 $\pm$ 0.041 \\
$\Gamma_{24}$ & 0.7207 $\pm$ 0.0077 & 0.1804 $\pm$ 0.0538 & 0.942 $\pm$ 0.041 & 0.421 $\pm$ 0.129  & -0.017 $\pm$ 0.054 & -0.019 $\pm$ 0.046   \\
$\Gamma_{3}$  & 0.7215 $\pm$ 0.0078 & 0.1596$\pm$ 0.0207 & 0.944 $\pm$ 0.038 & 0.389$\pm$ 0.119 & -0.004 $\pm$ 0.009 & - \\
$\Gamma_{4}$ & 0.7213$\pm$ 0.0077 & 0.1640 $\pm$ 0.0373  & 0.955 $\pm$ 0.031 &  0.423$\pm$ 0.144 & 0.002$\pm$ 0.011  & -  \\
$\Gamma_{5}$ & 0.7217 $\pm$ 0.0080 & 0.2080 $\pm$ 0.0854 & 0.967 $\pm$ 0.023 & 0.545 $\pm$ 0.213 & 0.014 $\pm$ 0.031  & -  \\
\hline
\end{tabular}%
    }
\caption{Best-fit parameters for the joint analysis Pantheon + CC + BAO + CMB. These results consider the priors in Table \ref{tab3} with the $h$ prior Gaussian.\label{tab4}}
\end{table}

\begin{table}
\centering
\begin{tabular}{ c  c  c  c  c  c  c }
\hline
Model & $h$ & $\Omega_{c}$ & $\alpha$ & $\beta$ & $\gamma$ & $\delta$ \\ \hline
$\Lambda \textrm{CDM}$& $0.6840 \pm 0.0054$ & $0.2541 \pm 0.0071$& -& -& - & -  \\
HRDE&$0.6823\pm0.0088$&$0.1780\pm0.0059$&-&$0.504\pm0.019$&-&-\\
MHRDE&$0.6889\pm0.0180$&$0.1807\pm0.0101$&$0.975\pm0.019$&$0.488\pm0.027$&-&-\\

$\Gamma_{11}$ & $0.6841 \pm 0.0181$ & $0.1729\pm 0.0145$ & $0.916 \pm 0.053 $ & $0.356 \pm 0.086$ & $0.037 \pm 0.026$  & -  \\
$\Gamma_{12}$ & $0.6831 \pm 0.0187$ & $0.1681 \pm 0.0257$ & $0.952 \pm 0.035$ & $0.416 \pm 0.111$ & -  & $0.014 \pm 0.028 $  \\
$\Gamma_{13}$ & $0.6842 \pm  0.0183$ & $0.1669 \pm 0.0189$ & $0.937 \pm 0.044$ & $ 0.384 \pm 0.104$ & $0.012 \pm 0.012$  &  $0.012 \pm 0.012$ \\
$\Gamma_{14}$ & $0.6832 \pm 0.0177$ & $0.1745 \pm 0.0171$ & $0.922 \pm 0.046$ & $0.369 \pm 0.079$  & $-0.023 \pm   0.040$& $0.047 \pm 0.033$  \\
$\Gamma_{21}$ & $0.6837 \pm 0.0183$ & $0.1755 \pm 0.0122$ & $0.927 \pm 0.050$ & $0.376 \pm 0.097$ & $-0.012 \pm 0.010$  & -\\
$\Gamma_{22}$ & $0.6838\pm0.0188$ & $ 0.1668 \pm 0.0190$ & $0.937 \pm 0.043$ & $0.384 \pm 0.104$ & -  & $ 0.012 \pm 0.012$ \\
$\Gamma_{23}$ & $ 0.6857 \pm  0.0184 $ & $0.2120 \pm 0.0297$ & $0.950 \pm 0.035$ & $ 0.475 \pm 0.030$ & $-0.042 \pm 0.038$  & $-0.042 \pm 0.038$ \\
$\Gamma_{24}$ & $0.6833 \pm 0.0183$ & $0.1888 \pm 0.4863$ & $0.938 \pm 0.045$ & $0.418 \pm 0.113$ & $-0.013 \pm 0.053$& $-0.021 \pm 0.046$   \\
$\Gamma_{3}$ & $0.6848 \pm 0.0185$ & $0.1688 \pm 0.0174$ & $0.941 \pm 0.041$ & $0.391 \pm 0.106$ & $-0.008 \pm 0.008$  & -\\
$\Gamma_{4}$ & $0.6855 \pm 0.0184$ & $0.1641 \pm  0.0290 $ & $0.951 \pm 0.034$ & $0.403\pm 0.121$ & $0.008 \pm 0.009$  & - \\
$\Gamma_{5}$ & $0.6872 \pm 0.0188$ & $0.182 \pm 0.076$ & $0.967 \pm 0.023$ & $0.483 \pm 0.193$ & $-0.002 \pm 0.029$  & - \\
\hline
\end{tabular}%

\caption{Best-fit parameters for the joint analysis Pantheon + CC + BAO + CMB. These results consider the priors in Table \ref{tab3} with the $h$ prior uniform.\label{tab5}}

\end{table}

\begin{table}[htb]
\centering
\begin{tabular}{ c | c  c  c | c c c}
\hline
\multirow{2}{*}{Model} & \multicolumn{3}{c}{Gaussian Prior} & \multicolumn{3}{|c}{Uniform Prior} \\
\cline{2-7}
 & $\ln{\mathcal{E}}$ & $\ln{B}$ & Interpretation & $\ln{\mathcal{E}}$ & $\ln{B}$ & Interpretation  \\ \hline
$\Lambda\text{CDM}$ & -540.172 $\pm$ 0.029 & & & -534.081 $\pm$ 0.007 & & \\
HRDE & -542.252 $\pm$ 0.014 &  -2.080 $\pm$ 0.032 & Weak & -537.466 $\pm$ 0.008  & -3.385 $\pm$ 0.011  & Moderate  \\

MHRDE & -540.563 $\pm$ 0.468 & -0.391 $\pm$ 0.469& Inconclusive & -540.611 $\pm$ 0.001 & -6.530 $\pm$ 0.012 & Strong  \\
$\Gamma_{11}$ & -541.522 $\pm$ 0.079 & -1.350 $\pm$ 0.084& Weak & -540.670 $\pm$ 0.022 & -6.589 $\pm$ 0.023 & Strong  \\
$\Gamma_{12}$ & -541.802 $\pm$ 0.035& -1.630$\pm$ 0.045 & Weak & -540.737$\pm$ 0.073 & -6.656 $\pm$ 0.073 & Strong   \\
$\Gamma_{13}$ & -542.395 $\pm$ 0.054 & -2.223 $\pm$ 0.061 & Weak & -541.321 $\pm$ 0.028 & -7.240 $\pm$ 0.029 & Strong \\
$\Gamma_{14}$ & -542.540 $\pm$ 0.049 & -2.368 $\pm$ 0.057 & Weak &  -541.322 $\pm$ 0.160 & -7.241 $\pm$ 0.160 & Strong  \\
$\Gamma_{21}$ & -542.677  $\pm$ 0.032   & -2.505 $\pm$ 0.043  & Moderate & -541.292 $\pm$ 0.130 & -7.211 $\pm$ 0.130 & Strong \\
$\Gamma_{22}$ & -542.367  $\pm$ 0.104   &  -2.195 $\pm$ 0.108  & Weak & -541.368 $\pm$  0.082 & -7.287 $\pm$ 0.082 & Strong \\
$\Gamma_{23}$ & -542.730 $\pm$ 0.176 & -2.558 $\pm$ 0.178 & Moderate & -541.476 $\pm$ 0.159 & -7.395 $\pm$ 0.159 & Strong  \\
$\Gamma_{24}$ & -541.401 $\pm$ 0.021 & -1.229 $\pm$ 0.036 & Weak & -540.538 $\pm$ 0.013 & -6.457 $\pm$ 0.015  & Strong  \\
$\Gamma_{3}$ & -542.970 $\pm$ 0.028 & -2.798 $\pm$ 0.040 & Moderate & -542.032 $\pm$ 0.021 & -7.951 $\pm$ 0.022 & Strong  \\
$\Gamma_4$ & -542.883 $\pm$ 0.059  & -2.711 $\pm$ 0.066 & Moderate & -541.975 $\pm$ 0.027 & -7.894 $\pm$ 0.028 & Strong \\
$\Gamma_5$ & -541.876 $\pm$ 0.052 & -1.740 $\pm$ 0.060 & Weak & -540.983 $\pm$ 0.114 & -6.902 $\pm$ 0.114 & Strong \\
\hline
\end{tabular}
\caption{Bayes' evidence (ln $\mathcal{E}$) and interpretation, for the joint analysis Pantheon + CC + BAO + CMB in the MHR-IDE models presented in Table \eqref{tab1}. Notice ln$B_i=$ln$\mathcal{E}_i-$ln$\mathcal{E}_{\Lambda \textrm{CDM}}<1$ favors the $\Lambda$CDM model. \label{tab6}}
\end{table}

\section{Final Remarks}\label{remarks}
We study a new class of modified holographic Ricci interacting dark energy models (MHR-IDE) that include new interactions in the dark sector. Given that no previous MHR-IDE models have been successful in fitting the data better than $\Lambda$CDM, we considered it interesting to conduct more exhaustive research to assess a wider range of interactions in the holographic context. This work can be considered as an extension of Ref. \cite{Cid:2020kpp}, with five new types of interactions revised.

In particular, five new MHR-IDE models have been studied. In section \ref{sec2} we present the models and analytical solutions. In section \ref{sec3} we perform the dynamical system analysis for most of the models. This analysis indicates that the radiation epoch is always modified for these models, but by considering $\alpha-2\beta\ll1$ and a small interaction, the modifications are ameliorated. A critical point corresponding to an attractor is present in each model and it represents an accelerated phase.

The studied models have been fitted using Bayesian inference techniques for a joint analysis of the dataset Pantheon + CC + BAO + CMB. We investigate whether these new MHR-IDE models are competitive against the $\Lambda$CDM model, in the framework of Bayesian comparison.

As an overall result, we find that the Bayesian comparison favors the $\Lambda$CDM model, irrespective of the considered interaction and independent of the prior assigned to the parameter $h$ (Gaussian or Uniform).

Notice that in our analysis, the MHRDE model presents inconclusive evidence in Table \ref{tab6} when a Gaussian prior in $h$ is used, in contrast with the strong evidence against this model when a Uniform prior is considered. This change in the evidence may be artificial and due to the tension between the CMB data used and the local measurement of $H_0$ we are using, certainly, more analysis is needed to firmly assess this conclusion.

Finally, we conclude that the new MHR-IDE models studied do not contribute to improving the Bayesian evidence or parameter estimation with respect to the already known holographic dark energy models \cite{Cid:2020kpp}. Most of the interacting scenarios revised in this work are compatible with a null interaction, and present certain correlation in the holographic parameters ($\alpha\approx2\beta$), accounting for the fact that the Bayesian analysis seems to indicate that MHR-IDE scenarios are compatible with the MHRDE model.\\

{\it Acknowledgements.} AC acknowledges the support of Vicerrectoría de Investigación y Postgrado of Universidad del Bío-Bío through grant no. 2120247 IF/R.


\bibliographystyle{ieeetr}
\bibliography{bibliography.bib}

\appendix

\section{Contour plots}\label{append}

\end{document}